\documentclass[12pt,a4paper]{article}%

\usepackage{lmodern}
\usepackage[T1]{fontenc}
\usepackage[utf8]{inputenc}
\usepackage{parskip}
\usepackage{amssymb}
\usepackage{amsfonts}
\usepackage{amsmath}
\usepackage{mathtools}
\usepackage{enumitem}
\usepackage{geometry}
\usepackage[authoryear, sort]{natbib}
\defcitealias{rdc:2019}{RDC, 2018, 2019}
\usepackage{multirow}
\usepackage{setspace}
\usepackage{booktabs}
\usepackage{graphicx}
\usepackage{adjustbox}
\usepackage[font=small,labelfont=sf]{caption}
 \usepackage[dvipsnames]{xcolor}
\usepackage{array}
\newcolumntype{L}{>{$}l<{$}}

\usepackage[colorlinks=false,
            urlcolor=blue,
            citecolor=blue,
            linkcolor=blue,
            pdfauthor={Alexander Mayer, Universita Ca' Foscari, Venice, Italy, and Dominik Wied, Uni Köln, Germany},
            pdftitle={Endogeneity Corrections in Binary Outcome Models with Nonlinear Transformations: Identification and Inference},
            bookmarks,
            bookmarksopen=true,
            bookmarksnumbered=true,
            pdfstartview={FitH},
            ]{hyperref}

\newcommand{\Tr}{{\textnormal{\textsf{T}}}}
\newcommand{\Ex}{\textnormal{\textsf{E}}}

\newcommand{\var}{\textnormal{\textsf{var}}}
\newcommand{\cov}{\textnormal{\textsf{cov}}}

\DeclareMathOperator*{\ssup}{\textnormal{\textsf{sup}}}
\DeclareMathOperator*{\llim}{\textnormal{\textsf{lim}}}
\DeclareMathOperator*{\plim}{\textnormal{\textsf{plim}}}

\DeclareMathOperator*{\lln}{\textnormal{\textsf{ln}}}
\DeclareMathOperator*{\mmax}{\textnormal{\textsf{max}}}

\usepackage{accents}

\newtheorem{remark}{Remark}
\newtheorem{assumption}{Assumption}

\newtheorem{proposition}{Proposition}

\begin{document}

\title{\vspace*{-2cm}Endogeneity Corrections in Binary Outcome Models with Nonlinear Transformations: Identification and Inference}

\author{Alexander Mayer\footnote{{\it Corresponding author}, email: \href{mailto:alexandersimon.mayer@unive.it}{alexandersimon.mayer@unive.it}. We would like to thank the associate editor and two referees for valuable comments and suggestions that helped to improve the paper. We are grateful to Florian Köhler for enabling access to the German administrative data from the \textit{Forschungsdatenzentren der Statistischen Ämter des Bundes und der Länder} as well as to Jörg Breitung, Rouven Haschka, Francis Vella, and Rafael Weißbach for helpful suggestions.}\\
{\it Università Ca' Foscari}\\
Venezia, Italy\\\vspace{-1em}
  \and
Dominik Wied\\
{\it University of Cologne}\\
Cologne, Germany}
\date{\today}
\maketitle
\thispagestyle{empty}
\vspace*{-.5cm}
\begin{abstract} \noindent 
For binary outcome models, an endogeneity correction based on nonlinear rank-based transformations is proposed. Identification without external instruments is achieved under one of two assumptions: either the endogenous regressor is a nonlinear function of one component of the error term, conditional on the exogenous regressors, or the dependence between the endogenous and exogenous regressors is nonlinear. Under these conditions, we prove consistency and asymptotic normality. Monte Carlo simulations and an application to German insolvency data illustrate the usefulness of the method. \\
  \vspace{1em}\noindent{\bf Keywords:} Control function, rank-based transformation, insolvency data
  \textbf{Words}:  7111
\end{abstract}

\newpage

\section{Introduction}

Estimating regression models in the presence of endogeneity without external instruments has become popular in econometrics (see, e.g. \citealp{lewbel:1997}, \citealp{rigobon:2003}, \citealp{ebbes:2005}, \citealp{kleinvel:09}, \citealp{dong:2010}, \citealp{esc:16}, \citealp{tran:22}, \citealp{lewetal:23}, \citealp{kiviet:23}, or  \citealp{gaowang:2023}) and in many applications in business and economics. Examples include, among others, empirical marketing science (see, e.g. \citealp{burmeister:15}, \citealp{bach:21}, or \citealp{zhang:24}), productivity analysis (see, e.g. \citealp{tran:2015} or \citealp{haschka:24a})  or  energy economics (see, e.g. \citealp{aloui:16}).

The starting point of this paper is the seminal work by \citet{guptapark:2012}, who propose such an estimator based on a specific copula assumption concerning the dependence between the endogenous regressor and the error term. \citet{yang:2022} and \cite{haschka:2022, haschka:24} provide extensions for the case that there is dependence between endogenous and exogenous regressors. \citet{bmw:24} consider an approach based on control functions and derive asymptotic properties for their estimator. A recent literature review can be found in \citet{parkgupta:2024}, see also \cite{becker:22}, \cite{papa:2022}, \cite{lien:2024}, or \cite{yietal:24}.

We fill a gap in the literature by considering binary outcome models in the presence of endogeneity without external instruments and with possibly nonlinear dependence between endogenous and exogenous regressors. Similar to the seminal work of \cite{rivers:88}, we identify the structural parameters using a control function approach. The crucial distinction is, however, that we do not require outside instruments. In particular, as a direct extension to \citet{bmw:24}, we propose to estimate these models with a nonparametrically generated control function to take potential endogeneity into account. This control function is derived from a rank-based transformation of the reduced-form residuals. In contrast to \citet{bmw:24}, however, we explicitly allow for nonlinear dependence between endogenous regressor and exogenous regressor in the first step, which is similar in nature to the approach taken by \cite{dong:2010} or \cite{esc:16}. In doing so, we allow for parametric, semi-parametric, and nonparametric estimation of the first stage. The so-obtained residuals are transformed to ranks, the ranks are transformed with the standard normal quantile function, and the resulting term is added as a control term to the regression equation.

This way, we obtain identification of the parameters, if the endogenous regressor is a nonlinear function of one component of the error term, conditional on the exogenous regressors. In this case, the transformation in the first step may be linear or nonlinear. Moreover, there is identification if the endogenous regressor is a linear function of the error term, but the dependence in the first step is nonlinear.

The possibility of linearity in the first step is an improvement over the existing literature. For example, \citet{dong:2010} considers a special case of our model, in which the transformation in the first step regression has to be nonlinear to achieve identification. In this case, no restriction on the dependence between the latent error terms is required. This is also similar to the identification strategy in \citet[Section 4.2]{esc:16}, who require nonlinearity of the first step. Moreover, our estimation strategy,
based on rank-based transformations, is inspired by the seminal work of \cite{guptapark:2012} and differs from the kernel-based estimators used, for example, in \cite{dong:2010} and \cite{esc:16}.
 
The difficulty of deriving an asymptotic theory stems from the nonparametrically generated control function. Using recent results from residual empirical processes theory (\citealp{zhao:2020} and \citealp{zhao:2022}) for (nonparametrically) estimated normal scores, we are able to show that the estimator is consistent and asymptotically normal. We do so by establishing sufficient high-level conditions that allow for parametric, semi-, and nonparametric estimation of the reduced form regression function. Similar to \cite{pagan:1984}, the sampling uncertainty of having to estimate the control function affects the sampling distribution of the estimator. We thus propose a bootstrap procedure to take it into account.

A simulation study yields numerical evidence and an empirical application on German insolvency data illustrates the usefulness of our estimator. Here, we model the probability of the start of an insolvency case as a function of the recent company growth. We use a unique administrative dataset from the German \textit{Forschungsdatenzentren der Statistischen Ämter des Bundes und der Länder} \citepalias{rdc:2019} which contains over one million companies in the year 2018 and 2019. Correcting for potential endogeneity yields substantively different results, which are robust to controls and are in line with other results in the literature.

The remainder of this paper is organised as follows. Section \ref{sec:model} introduces the model, the (limited) maximum likelihood estimator, and discusses identification. We lay out assumptions, discuss consistency and asymptotic normality of
the estimator as well as inferential methods in Section \ref{sec:asymp}.  Section \ref{sec:mc} contains a Monte Carlo study, while the empirical application is presented in Section \ref{sec:emp}.

\section{Model, Identification, and Estimator}\label{sec:model}

Following \cite{rivers:88},  we consider the following structural model
\begin{align}\label{eq:model} 
Y  = 1\{\alpha^\Tr Z + \beta  D + U \geq 0\},
\end{align}
where \(Z\) is a \(k \times 1\) vector of exogenous regressors (including a constant) and \(D\) is a scalar endogenous regressor correlated with the error term \(U\). Let us further assume that the endogenous variate and the error can be decomposed as
\begin{align}\label{eq:decomp}
D =  \pi(Z)+V \quad \text{ and } \quad U = \rho m(V) + E,
\end{align}
where \(V\) (continuous) and \(E\) are mean-zero error terms independent of $Z$ and $Z$ and $D$, respectively. The functions $\pi(\cdot)$ and $m(\cdot)$ are in general unknown. Thus, unless $\rho=0$, $D$ is endogenous due to the presence of the term $m(V)$.

One object of interest could be for example the average structural function (ASF). Assuming $E/\sigma \sim F$, for some symmetric cumulative distribution function (cdf) $F$, we get
\(
\Ex[Y \mid Z,D,V]=  F(\sigma^{-1}(\alpha^\Tr Z+\beta D+\rho m(V))),
\)
because $E$ is independent of $Z$, $D$, and thus also of $V = D-\pi(Z)$. For example, if $F$ is the standard normal cdf $\Phi$, say, and $m(V) \sim \Phi$, then 
the ASF given by
\begin{align}\label{eq:ASF}
{\sf ASF}(x) \coloneqq \Phi\left(\frac{\alpha'z+  \beta d}{\sqrt{\sigma^2+\rho^2}}\right), \quad x = (z^\Tr,d)^\Tr \in \mathbb{R}^{k+1}.
\end{align}
To make these objects operational, an estimator of the unknown parameters is needed.

Our estimator is based on the following fundamental identification assumption.

\begin{assumption} \label{ass:ident} $m(V) \sim H$ and $V \sim G$, where are mean-zero cdf's with continuous densities. Moreover, (1) $m$ is strictly monotone, (2) either $G$ or $H$ are strictly monotone, and (3) one of the following holds true
\begin{itemize}
    \item[\textnormal{($a$)}]  $v \mapsto m(v)$ is a  nonlinear function.
    \item[\textnormal{($b$)}]  $z \mapsto \pi(z)$ is a nonlinear function.
\end{itemize}   
\end{assumption}

If the marginal distribution $G$ of the reduced-form error term $V$ were known and Assumption \ref{ass:ident} ($a$) would hold, then $m(V) = H^{-1}(G(V)) \eqqcolon \eta$ almost surely (see the proof of \citealp[Theorem 2.1]{bmw:24}), and we could (up to $\sigma$) identify  $\theta = (\gamma^\Tr,\rho)^\Tr$, $\gamma = (\alpha^\Tr,\beta)^\Tr$, by augmenting the model using the so-called {\it `control function'} $\eta$, i.e.\  $Y = 1\{\alpha^\Tr Z + \beta D + \rho\eta + E\geq0\}$. This is essentially the identification strategy of the popular copula-based endogeneity correction of \cite{guptapark:2012} and related approaches; see \citet{parkgupta:2024} for a recent review. \footnote{ Monotonicity of \( m \), together with that of either \( H \) or \( G \), implies strict monotonicity of the triple \( (m, H, G) \), which is essential to pointwise identify the control function via \( m(V) = H^{-1}(G(V)) \), and, in conjunction with Assumption \ref{ass:ident} (a), to ensure \( G \neq H \). If any of the functions \( m \), \( G \), or \( H \) fail to be strictly monotonic, then the identity \( m(V) = H^{-1}(G(V)) \) may hold only in distribution, not almost surely. Moreover, if \( G = H \), then strict monotonicity implies \( G(v) = H(v)={\sf P}(V \leq m^{-1}(v))=G(m^{-1}(v))  \), and by injectivity of \( G \), it follows that \( m = {\sf id} \), contradicting Assumption~\ref{ass:ident} (a).} If, however, $G=H$, then part ($a$) is violated and identification breaks down as $(Z,D,\eta) = (Z, \pi(Z)+V, V)$ are perfectly collinear unless $\pi(\cdot)$ is nonlinear. In other words, non-linearity of the first stage provides an additional source of identification and yields a `{\it robustification}' against violations of part ($a$) (i.e. $G=H$). This aspect of our identification strategy is novel relative to the earlier work cited above and applies also to the linear models considered in \cite{bmw:24} and thus also to several specifications derived from the seminal approach of \cite{guptapark:2012}.

\begin{remark}
Note that in our exposition, we considered the case of a single endogenous regressor. As noted by \textnormal{\citet[Remark 2.1]{bmw:24}}, it is, in principle, possible to extend this framework to multiple endogenous regressors by additively incorporating the rank-based control variables for each regressor. A more challenging extension would be to allow for a noncontinuous treatment variable $D$ (e.g. binary). \textnormal{\citet{lewbel:2018}} addresses this in a different model under strong assumptions. One possible approach is to define a latent variable $\tilde D = 1\{ D \geq 0\}$, where $D = \pi(Z)+V$ follows the specification above. We leave this for further research.
\end{remark}

In practice, $G$ is typically unknown and has to be estimated from a sample $\mathcal{S}_n \coloneqq \{X_i,Y_i\}_{i=1}^n$, say, which is assumed to be an {\sf IID} sample independently drawn from $X\coloneqq (Z^\Tr,D)^\Tr$ and $Y$. More specifically, we could estimate $G$ in a first step nonparametrically and construct
\begin{align}\label{eq:etatil}
 \tilde\eta_{i,n} \coloneqq H^{-1}(\tilde G_{n}(V_{i})) \quad  \tilde G_n(t) \coloneqq \frac1{n+1}\sum_{i=1}^n1\{V_{i}\leq t\}, \quad t\in \mathbb{R}.
\end{align}
Note that $\tilde G_n(V_i)$ is just the relative empirical rank, i.e.\ the rank of $V_i$ among $\{V_1,\dots,V_n\}$ divided by $n+1$. Since also $\pi(\cdot)$ is typically unknown, we could estimate the regression function in a preliminary step to obtain the residuals $V_{i,n} = D_i  - \pi_n(Z_i)$ for some estimator $\pi_n(\cdot)$ constructed from $\mathcal{S}_n$ so that
\begin{align}\label{eq:eta}
\eta_{i,n} \coloneqq H^{-1}(G_{n}(V_{i,n})), \quad G_n(t) \coloneqq \frac1{n+1}\sum_{i=1}^n1\{V_{i,n}\leq t\}.
\end{align}
Our instrument-free approach comes at the cost of having to specify $H$, the cdf of the source of endogeneity $m(V)$. In principle, different choices are possible. Here, we follow the literature (see \citealp{parkgupta:2024} and the references therein), and make the following normality assumption as this allows us to leverage corresponding theoretical results for normal scores (\citealp{zhao:2020,zhao:2022}). 
\begin{assumption}\label{ass:H}
    $H = \Phi$, where $\Phi$ is the standard normal cdf.
\end{assumption}

Finally, in order to derive our estimator a link function, i.e.\ the cdf of the innovation $E$ has to be fixed. The following conditions restrict the link function and impose restrictions on $V$ and $E$.

\begin{assumption}\label{ass:F}
    ($i$) For some $\sigma \in (0,\infty)$, assume that $E/\sigma$ has cdf $F$. $F$ has derivative $f$ and second-order derivative $f'$, and $0 < F(x) < 1$ and $f(x)> 0$ for every $x$. ($ii$) $E$ is mean-zero and independent of $Z$ and $D$. ($iii$) $V$ is mean-zero with finite variance and independent of $Z$.
\end{assumption}

Part ($i$) is a common assumption in the literature and identical to \citet[Assumption 9.2.1]{amemiya:1985}. Popular choices for $F$, that satisfy part ($i$), include the logistic (i.e.\ $F(z) = \Lambda(z) \coloneqq {\sf exp}(z)/(1+{\sf exp}(z))$) or the standard normal distribution (i.e.\ $F = \Phi$), giving rise to a logit and a probit specification, respectively. Part ($ii$) implies that $E$ is independent of $Z$, $D$, and $V$, while part ($iii$) requires independence between $V$ and $Z$.

Now, set $X \coloneqq (Z^\Tr,D)^\Tr$ and define the log-likelihood contribution
\begin{equation}\label{eq:lik}
\ell(\theta; Y,X,\eta) \coloneqq Y  \lln(F(X^\Tr \gamma +  \rho \eta)) + (1-Y)\lln(1-F(X^\Tr \gamma +  \rho \eta)),
\end{equation}
where we use the normalization $\sigma = 1$. Our limited information maximum likelihood estimator $\theta_n$, say, of the true parameter vector $\theta_0$ is an extremum estimator defined as a solution (if it exists) of
$$
\frac{\partial}{\partial \theta}{\cal L}_n(\theta) = 0, \quad {\cal L}_n(\theta) \coloneqq \frac1{n}\sum_{i=1}^n\ell(\theta;Y_i,X_i,\eta_{i,n}).
$$
This is similar to \cite{rivers:88} in that we use a control function to cope with endogeneity in a binary response model, with the distinction, however, that our approach does not require outside instruments.  

\section{Asymptotic Theory and Inference}\label{sec:asymp}

\subsection{Consistency}

A crucial tool to derive the asymptotic properties of $\theta_n$ is the following high-level assumption on the difference between the infeasible control function $\tilde\eta_{i,n}$ and its feasible counterpart $\eta_{i,n}$, both defined in Eqs. \eqref{eq:etatil} and \eqref{eq:eta}, respectively.

\begin{assumption}\label{ass:ranks} Set $U \coloneqq G(V)$ and let $\mathbb{E}_n[r_n(Y,X,U)] = \Ex[r_n(Y,X,U) \mid \mathcal{S}_n]$ be the expectation conditional on $\mathcal{S}_n = \{X_i,Y_i\}_{i=1}^n$ for some measurable function $r_n$, possibly depending on ${\cal S}_n$. Then,
    \begin{align}\tag{$i$}
    \eta_{i,n}-\tilde\eta_{i,n} = -\kappa(U_i) (\pi_n(Z_i)-\mathbb{E}_n[ \pi_n(Z)]-\pi(Z_i)+\Ex[\pi(Z)]) + R_{i,n},  
    \end{align}
    with $\mmax\limits_{i\leq 1 \leq n} |R_{i,n}| = o_p(n^{-1/2})$ and 
    \begin{align}\tag{$ii$}
    \frac1{n}\sum_{i=1}^n[\kappa(U_i)(\pi_n(Z_i)-\mathbb{E}_n[ \pi_n(Z)]-\pi(Z_i)+\Ex[\pi(Z)])]^2 = o_p(1)
     \end{align}
     for some square integrable function $\kappa: [0,1] \mapsto \mathbb R$.
\end{assumption}

    If the regression function is linear $\pi(z) = z^\Tr\delta$, and the $k \times 1$ vector $\delta$ is estimated by the OLS estimator $\delta_n$, say, then as discussed in \cite{bmw:24}, it follows from \cite{zhao:2020} that Assumption \ref{ass:ranks} is satisfied for 
     \[
    \eta_{i,n}-\tilde\eta_{i,n} = -\kappa(U_i)(\delta_n-\delta)^\Tr(Z_i-\Ex[Z]) + R_{i,n}, \quad \kappa(u) = \frac{g(G^{-1}(u))}{\phi(\Phi^{-1}(u))},
    \]
    with $\mmax_i |R_{i,n}| = o_p(n^{-1/2})$.   If, in addition, $\Ex[\Vert \kappa(U)Z\Vert^2] < \infty$, 
    then  also Assumption \ref{ass:ranks} ($ii$) will be satisfied. If $\pi(\cdot)$ is a smooth function estimated nonparametrically using common kernel-based estimation techniques, then Assumption \ref{ass:ranks} holds for $\kappa(\cdot)$ defined above under standard regularity conditions as demonstrated in \cite{zhao:2022}; for part ($ii$) see, e.g. \cite{hardle:88}.

To derive consistency, we show that ${\cal L}_n(\theta)$ is, uniformly in $\theta \in \Theta \subseteq \mathbb{R}^{k+2}$, close to the infeasible objective function ${\cal L}_{0,n}(\theta) = \frac{1}{n}\sum_{i=1}^{n} \ell(\theta;Y_i, X_i,\eta_i)$. To do so, we have to impose a mild regularity condition to control small perturbations in a neighbourhood around ${\cal L}_{0,n}(\theta)$. More specifically, define
\[
\psi(\theta;Y,X,\eta) \coloneqq \frac{Y-F(X^\Tr\gamma + \rho \eta)}{F(X^\Tr\gamma + \rho \eta)(1-F(X^\Tr\gamma + \rho \eta))}f(X^\Tr\gamma + \rho \eta),
\]
and note that $\rho\psi(\cdot;\cdot,\cdot,t)  = \partial \ell(\cdot; \cdot,\cdot,t)/\partial t$. For example, in the probit case ($F=\Phi$), we obtain
\[
\psi(\theta;Y,X,\eta) = \frac{Y-\Phi(X^\Tr\gamma + \rho \eta)\phi(X^\Tr\gamma + \rho \eta)}{(1-\Phi(X^\Tr\gamma + \rho \eta))\Phi(X^\Tr\gamma + \rho \eta)},
\]
while for a logit specification ($F = \Lambda$), we get $\psi(\theta;Y,X,\eta) = (Y-\Lambda(X^\Tr\gamma + \rho \eta))$.
 
\begin{assumption}\label{ass:psi} For any $b_n = o(1)$ as $n \rightarrow \infty$,
\[
\ssup\limits_{\theta \in \Theta}\frac{1}{n}\sum_{i=1}^{n}\ssup_{\{t_i: |t_i-\eta_i|\leq b_n\}}\psi^2(\theta;Y_i,X_i,t_i)= O_p(1).
\]
\end{assumption}

Following the discussion in \cite{baing:2008}, Assumption \ref{ass:psi} can be shown to be satisfied for logit or probit specifications of the link function.

\begin{proposition}\label{prop:consistency} Suppose Assumptions \ref{ass:ident}-\ref{ass:psi} are satisfied. If $\theta_0$ is contained in a open subset of $\mathbb R^{k+2}$ and uniquely minimizes  $\plim\limits_{n\rightarrow \infty} {\cal L}_{0,n}(\theta)$, then $\Vert \theta_n-\theta_0\Vert = o_p(1).$
\end{proposition}

\subsection{Limiting distribution}

Having established weak consistency of $\theta_n$, we now turn to the question of its asymptotic distribution. In a first step, the asymptotic normality of the score vector evaluated at the true value $\theta_0$ is established. To achieve this, we assume, similar to \citet[Assumption 8]{rothe:09}, that the feasible score function satisfies a linear representation. In particular, define the $(k + 2) \times 1$ score vector
$s(t;\cdot,\cdot,\cdot) \coloneqq \partial \ell(t;\cdot, \cdot,\cdot)/\partial t$, where
\begin{equation}
s(\theta;Y,X,\eta) \coloneqq   W  \psi(\theta;Y,X,\eta), \quad W \coloneqq (X^\Tr, \eta)^\Tr.
\end{equation}
Let us also define the first derivative of $\psi$ with respect to the last argument $\rho\dot{\psi}(\cdot;\cdot,\cdot,t) =\partial \psi(\cdot; \cdot,\cdot,t)/\partial t$ given by
\begin{equation}
\begin{split}
\dot{\psi}(\theta;Y,X,\eta) \coloneqq  \,&  \frac{Y-F(X^\Tr\gamma + \rho \eta)}{F(X^\Tr\gamma + \rho \eta)(1-F(X^\Tr\gamma + \rho \eta))}f'(X^\Tr\gamma + \rho \eta) \\
\,& -   \left[\frac{Y-F(X^\Tr\gamma + \rho \eta)}{F(X^\Tr\gamma + \rho \eta)(1-F(X^\Tr\gamma + \rho \eta))}f(X^\Tr\gamma + \rho \eta) \right]^2
 \end{split} 
\end{equation}
and set
$S(\theta;Y,X,\eta) \coloneqq  W\dot\psi(\theta;Y,X,\eta)$,  $S_0(Y,X,\eta) \coloneqq W  \dot\psi_0(Y,X,\eta),$
with $\dot\psi_0(Y,X,\eta) \coloneqq \dot\psi(\theta_0;Y,X,\eta).$\footnote{To illustrate, note that for a probit specification we get
\begin{equation}\nonumber
\begin{split}
\dot{\psi}(\theta;Y,X,\eta) = \,& -Y\lambda(X^\Tr\gamma + \rho \eta)(X^\Tr\gamma + \rho \eta+\lambda(X^\Tr\gamma + \rho \eta))\\
\,&\quad-(1-Y)\lambda(-X^\Tr\gamma - \rho \eta)(-X^\Tr\gamma - \rho \eta+\lambda(-X^\Tr\gamma - \rho \eta))    
 \end{split} 
\end{equation}
where $\lambda(x) \coloneqq \phi(x)/\Phi^{-1}(x)$,while for a logit model
\[
\dot{\psi}(\theta;Y,X,\eta) =  -\Lambda(X^\Tr\gamma + \rho \eta)(1- \Lambda(X^\Tr\gamma + \rho \eta)).
\]}
We then use the following linear representation:

\begin{assumption}\label{ass:linrep}
For some function $q(\cdot)$ such that $\Ex[\Vert q(Z)\Vert^2] < \infty$ and $\Ex[q(Z)q(Z)^\Tr]$ positive definite, it holds
\begin{equation}\nonumber
\begin{split}
     \frac1{\sqrt{n}} \sum_{i=1}^n S_0(Y_i,X_i,\eta_i)\kappa(U_i)(\pi_n(Z_i)-&\mathbb{E}_n[\pi_n(Z)]-\pi(Z_i)+\Ex[\pi(Z)]) \\
     \,& =  \frac1{\sqrt{n}} \sum_{i=1}^n V_iq(Z_i) + o_p(1).
\end{split}
\end{equation}
\end{assumption}
    This assumption warrants some discussion. First, suppose that $\psi(Z) = Z^\Tr\delta$, and $\delta$ is estimated by the OLS estimator $\delta_n$, then $\pi_n(Z_i)-\mathbb{E}_n[\pi_n(Z)]-\pi(Z_i)+\Ex[\pi(Z)] = (Z_i-\Ex[Z])^\Tr(\delta_n-\delta_0)$ and Assumption \ref{ass:linrep} is satisfied with
    \(
    h(Z_i) \coloneqq \cov[S_0(Y,X,\eta)\kappa(U),Z](\Ex[ZZ^\Tr])^{-1}(Z_i-\Ex[Z]).
    \)
    If, on the other hand, $\pi_n(Z)$ is e.g. the local linear kernel estimator, then Assumption \ref{ass:linrep} holds with
    \(
    h(Z_i) \coloneqq \Ex[S_0(Y_i,X_i,\eta_i)\kappa(U_i) \mid Z_i]-\Ex[S_0(Y,X,\eta)\kappa(U)].
    \)
  To see this, note that the left-hand side of the expression in Assumption \ref{ass:linrep} can be written as
\begin{equation}
\begin{split}
 \mathbb{G}_n(S_0(\cdot)\kappa(\cdot) \pi_n(\cdot))-&\mathbb{G}_n(S_0(\cdot) \kappa(\cdot)\pi(\cdot)) \\
\,& - \frac1{\sqrt{n}} \sum_{i=1}^n \mathbb{E}_n[S_0(Y_i,X_i,U_i)\kappa(U_i)(\pi_n(Z)-\pi(Z))]  \\
\,& + \sqrt{n}\mathbb{E}_n[S_0(Y,X,U)\kappa(U)(\pi_n(Z)-\pi(Z))],
\end{split} 
\end{equation}
where
\(
\mathbb{G}_n(r_n(\cdot)) \coloneqq \frac1{\sqrt{n}}\sum_{i=1}^n(r_n(X_i)-\mathbb{E}_n[r_n(X)]),\) for some function $r_n$ that might depend on ${\cal S}_n$. By stochastic equicontinuity (for primitive sufficient conditions see \citealp{escanciano2014uniform}), $\mathbb{G}_n(S_0(\cdot)\kappa(\cdot) \pi_n(\cdot))-\mathbb{G}_n(S_0(\cdot) \kappa(\cdot)\pi(\cdot)) = o_p(1)$, while
\[
 \frac1{\sqrt{n}} \sum_{i=1}^n \mathbb{E}_n[S_0(Y_i,X_i,U_i)\kappa(U_i)(\pi_n(Z)-\pi(Z))] =  \frac1{\sqrt{n}} \sum_{i=1}^n \Ex[S_0(Y,X,U)\kappa(U)]V_i + o_p(1),
\]
because, by the LLN, $n^{-1}\sum_{i=1}^nS_0(Y_i,X_i,U_i)\kappa(U_i) = \Ex[S_0(Y,X,U)\kappa(U)]+o_p(1)$ and
\(
\mathbb{E}_n[\pi_n(Z)-\pi(Z)] = \int (\pi_n-\pi)(z)f_Z(z)dz =  \frac1{n}\sum_{j=1}^n V_j + o_p(1).   
\)
Thus, Assumption \ref{ass:linrep} holds.

\begin{proposition}\label{prop:score} Suppose the conditions of Proposition \ref{prop:consistency} are met, Assumption \ref{ass:linrep} holds, and $\Omega_1 \coloneqq \Ex[s_0(y,X,\eta)(s_0(y,X,\eta))^\Tr]$ is positive definite. Then
    $$\frac1{\sqrt{n}}\sum_{i=1}^n s_0(Y_i,X_i,\eta_{i,n}) \rightarrow_d \mathcal{N}(0,A), \qquad  A \coloneqq \Omega_1+\rho^2(\Omega_2+\Omega_3),$$
where 
    \[\normalfont
    [\Omega_2]_{i,j} = \int_0^1\int_0^1 h_i(u)h_j(v)(\textsf{min}(u,v)-uv) {\sf d}u{\sf d}v, \quad h(u) \coloneqq \frac{\Ex[S_0(Y,X,\Phi^{-1}(U)) \mid U = u]}{\phi(\Phi^{-1}(u))}
    \]
for $i,j \in \{ 1,\dots,k+2\}$, while
\(
\Omega_3\coloneqq \var[V]\Ex[q(Z)(q(Z))^\Tr].
\)
\end{proposition}

Note that $\Omega_2 = 0$ if $G(\cdot)$ is known, $\Omega_3 = 0$ if $\pi(\cdot)$ is known, and $A = \Omega_1$ if $D$ is exogenous. 

In order to derive the limiting distribution of $\theta_n$, we have to make sure that the Hessian
\[
H(\theta;Y,X,\eta) \coloneqq \frac{\partial}{\partial \theta} s(\theta;Y,X,\eta) = -WW'\dot\psi(\theta;Y,X,\eta)
\]
obeys a uniform LLN. The following assumption is similar to \citet[Assumption M3]{baing:2008} and, using their arguments, can be shown to hold for logit and probit specifications.

\begin{assumption}\label{ass:unif}
\begin{equation}\nonumber
\begin{split}
 \ssup\limits_{\bar\theta_n: |\bar\theta_n-\theta_0|=o(1)}\frac1{n} \sum_{i=1}^n \ssup\limits_{\bar\eta_{i,n}: |\bar\eta_{i,n}-\eta_i|=o(1)}\left\Vert \frac{\partial^2}{\partial r\partial t}s(r;Y,X,t) \Bigg\vert_{t = \bar\eta_{i,n},r=\bar\theta_n}\right\Vert^2 = \,&O_p(1) \\ 
  \ssup\limits_{\bar\theta_n: |\bar\theta_n-\theta_0|=o(1)}\frac1{n} \sum_{i=1}^n \ssup\limits_{\bar\eta_{i,n}: |\bar\eta_{i,n}-\eta_i|=o(1)}\left\Vert \frac{\partial^2}{\partial t^2}s(\bar\theta_n;Y,X,t) \Bigg\vert_{t = \bar\eta_{i,n}}\right\Vert^2 = \,&O_p(1) \\
   \ssup\limits_{\bar\theta_n: |\bar\theta_n-\theta_0|=o(1)}\frac1{n} \sum_{i=1}^n \ssup\limits_{\bar\eta_{i,n}: |\bar\eta_{i,n}-\eta_i|=o(1)}\left\Vert \frac{\partial^2}{\partial r\partial r^\Tr}s(r;Y,X,\bar\eta_{i,n}) \Bigg\vert_{r=\bar\theta_n}\right\Vert^2 = \,&O_p(1)
\end{split} 
\end{equation}
\end{assumption}

\begin{proposition}\label{prop:normal} Suppose the conditions of Proposition \ref{prop:score} are met and Assumption \ref{ass:unif} holds. Then
    $$\sqrt{n}(\theta_n-\theta_0) \rightarrow_d \mathcal{N}(0,\Sigma), \quad \Sigma \coloneqq  \Omega_1^{-1}A \Omega_1^{-1}.$$
\end{proposition}

\subsection{Inference}

As Proposition \ref{prop:normal} reveals, the limiting distribution depends on unknown nuisance parameters. An important special case concerns hypotheses that contain the restriction of no endogeneity (i.e.\ $\rho = 0$). In this case we can use common textbook standard errors, as $\sqrt{n}(\theta_n-\theta_0) \rightarrow_d \mathcal{N}(0,\Omega_1^{-1})$, where $\Omega_1$ is  consistently estimated using standard approaches (\citealp[Section 4.5]{amemiya:1985}).

For more general hypotheses, we propose the following pairs bootstrap: Draw $(Y_{b,1},X^\Tr_{b,1})^\Tr$,$\dots$,$(Y_{b,n},X^\Tr_{b,n})^\Tr$ with replacement from the empirical distribution of the original data ${\cal S}_n$ and define, analogously to \(\theta_n\), \(\theta_{n,b}\) based on the bootstrap data. We can then construct bootstrap standard errors via
$\Sigma_{n,B} \coloneqq \frac{n}{B}\sum_{b=1}^B (\theta_{n,b}-\theta_n)(\theta_{n,b}-\theta_n)^\Tr$. Following the discussion surrounding \citet[Corrolary 1]{bmw:24}, consistency of $\Sigma_{n,B}$ conditionally on the original data ${\mathcal S}_n$ (as $n$ and $B$ diverge) follows if we assume that $\sqrt{n}(\theta_n-\theta_0)$ possesses uniformly integrable second moments. 

If interest lies in other functionals, for example, the ASF introduced in Eq. \eqref{eq:ASF}, one could apply the delta-method in conjunction with $\Sigma_{n,B}$ (\citealp[Section 15]{woold:2010}). To fix ideas, consider the probit-specification (i.e. $F = \Phi$), then we can estimate the ASF via
\[
{\sf ASF}_{n}(x) \coloneqq \Phi\left(\frac{\theta_{n,\alpha}^\Tr z+ \theta_{n,\beta} d}{\sqrt {1+\theta_{n,\rho}^2}}\right)
\]
for some $x = (z^\Tr, d)^\Tr \in \mathbb{R}^{k+1}$ and      the partition $\theta_n = (\theta_{n,\alpha}^\Tr,\theta_{n,\beta},\theta_{n,\rho})^\Tr$ corresponds to $W$. An estimator of the asymptotic variance is then given the sandwich form $(\nabla_\theta {\sf ASF}(x)\vert_{\theta = \theta_n})^\Tr \Sigma_{n,B} \nabla_\theta {\sf ASF}(x)\vert_{\theta = \theta_n}$, where $\nabla_\theta {\sf ASF}(x)$ is the $(k+2) \times 1$ gradient vector.

\subsection{Relaxing Assumptions 2 and 3}

{\it Unknown distribution of the endogeneity.} Instead of requiring that $m(V) \sim  H=H_0$ is known, one could assume that the unknown $H_0$ belongs to a class of parametric distributions ${\mathcal H}$, say, for which $H_0 \coloneqq H(\lambda)$ is known up to  a finite dimensional parameter $\lambda = \lambda_0 \in {\sf int}(\Lambda)$, $\Lambda \in \mathbb{R}^l$. Our estimator of $(\theta_0^\Tr,\lambda_0^\Tr)^\Tr$ then maximizes the following objective function:
\begin{align*}
 {\cal L}^\dagger_n(\theta,\lambda) \coloneqq \,&\frac1{n}\sum_{i=1}^n (Y_i  \lln(F(X_i^\Tr \gamma +  \rho \eta_{i,n}(\lambda)) + (1-Y_i)\lln(1-F(X_i^\Tr \gamma +  \rho \eta_{i,n}(\lambda))),
\end{align*}
where $\eta_{i,n}(\lambda) \coloneqq H^{-1}(G_n(V_{i,n});\lambda)$. Given that $H^{-1}(\lambda)$ enters the objective function, it would be desirable to specify $H(\lambda)$ such that the inverses have closed-form representations. One such computationally appealing yet flexible choice of ${\cal H}$ could be the class of asymmetric distributions considered by \cite{gijbels:2019}. A special case is the two-piece skew-normal distribution of \cite{mudhut:00} for which 
\begin{equation}
\begin{split}
H^{-1}(u;\lambda) \coloneqq  (1\ +&\ \lambda)\  \Phi^{-1}\left(\frac{u}{1+\lambda}\right) {1}\{u < (\lambda+1)/2\}\\
& + (1-\lambda)\ \Phi^{-1}\left(\frac{u-\lambda}{1-\lambda}\right) {1}\{u \geq (\lambda+1)/2\}, \; -1 < \lambda < 1.
\end{split}
\end{equation}
Adopting this specification, Assumption \ref{ass:H} can be empirically tested via $H_0$: $\lambda = 0$. Although, in principle, the properties of the estimator could be investigated by leveraging the results developed here and the likelihood theory obtained by \cite{gijbels:2019}, a theoretical treatment is well beyond the scope of the current paper. 

{\it Unknown distribution of the innovation.} In case the true link function $F=F_0$  is unknown, one could replace $F_0$ in Eq. \eqref{eq:lik} with a nonparametric estimate so that our estimator $\theta^\ddagger_n$, say, maximizes
\begin{align*}
 {\cal L}^\ddagger_n(\theta) \coloneqq \,&\frac1{n}\sum_{i=1}^n \tau_{i} (Y_i  \lln(F_{n}(X_i^\Tr \gamma +  \rho \eta_{i,n})) + (1-Y_i)\lln(1-F_{n}(X_i^\Tr \gamma +  \rho \eta_{i,n}))),
\end{align*}
where $\tau_i \coloneqq 1\{ (X_i,\eta_{i,n}) \in {\cal X} \}$ is a trimming function for a compact set ${\cal X}$ and, for a given $\theta$, $F_{n}$ is the Nadaraya-Watson estimator of $F_0$, i.e. a nonparametric kernel regression of $Y_i$ on $X_i^\Tr\gamma + \rho\eta_{i,n}$, $i \in \{1,\dots,n\}$. The dependence of $F_n$ on $\theta$ is implicitly understood for ease of notation. This is idea is similar to the approach proposed initially by \cite{klein:93} and then further extended by, among others,  \cite{blundell:2004} and \cite{rothe:09}.  Adapting the arguments of \citet[Section 4]{rothe:09}, it might be possible to show that $\theta^\ddagger_n$ is $\sqrt{n}$-consistent with asymptomatic Gaussian limiting distribution; a conjecture supported by the finite sample evidence of the following section. Finally, a nonparametric estimator of the ASF in Eq. \eqref{eq:ASF} can be obtained via
\[
\widetilde{\sf ASF}_n(x) \coloneqq\frac1{n}\sum_{i=1}^n F_n(x^\Tr \theta^\dagger_{n,\gamma} + \theta^\dagger_{n,\rho}\eta_{i,n}), \quad x \in \mathbb{R}^{k+1}.
\]

\section{Monte Carlo Simulation}\label{sec:mc}

The Monte Carlo design is based on Eq. \eqref{eq:model}, i.e.\ $Y = 1\{\alpha_0 + \alpha_1 Z + \beta D+ U>0\}$, where $Z \sim \Phi$ and $D=\pi(Z)+V$. We consider a probit specification, where $E \sim \Phi$ in $U = \rho m(V)+E$. The assumption that $m(V) \sim \Phi$ is maintained throughout, while the distribution $G$ of the reduced form error $V$ is $G = \Phi$ or $G = {\sf Gamma}(2,2)$, respectively. We distinguish between a linear ($\pi(z) = z$) and a non-linear ($\pi(z) = z^2$) specification of the reduced form. It is apparent from the discussion of Assumption \ref{ass:ident} that identification breaks down if $G$ is normal and $\pi(\cdot)$ is linear.

\setlength{\tabcolsep}{1.4pt}
\begin{table} 
\caption{Monte Carlo results for $n=500$/$\rho = 0.50$}
\label{tab:mc1}
\vspace*{-.25cm}
\begin{adjustbox}{max width=\textwidth}
\begin{tabular}{lrccccccccccccccccccc} \toprule \\[-.5cm]  \hline\\[-.2cm]
&&\multicolumn{9}{c}{$\pi(z) = z$}&&\multicolumn{9}{c}{$\pi(z) =z^2$}\\[-.1cm]
\cmidrule(r){3-11} \cmidrule(r){13-21} \\[-.5cm]
&&\multicolumn{4}{c}{$V\sim\Phi$} & &  \multicolumn{4}{c}{$V\sim{\sf Gamma}(2,2)$} &&\multicolumn{4}{c}{$V\sim\Phi$} & &  \multicolumn{4}{c}{$V\sim{\sf Gamma}(2,2)$} \\[-.1cm]
 \cmidrule(r){3-7} \cmidrule(r){8-11} \cmidrule(r){13-17} \cmidrule(r){18-21} \\[-.5cm]
&  & \sf mean  &   \sf std   &  \sf rmse  & \sf size  &&  \sf mean  &   \sf std   &  \sf rmse  & \sf size & & \sf mean  &   \sf std   &  \sf rmse  & \sf size  &&  \sf mean  &   \sf std   &  \sf rmse  & \sf size\\  \cmidrule(r){2-21}
\multirow{4}{*}{\rotatebox[origin=c]{90}{$\alpha_0 = 0.50$}} 	& {\sf ML}&		0.5067	&	0.0887	&	0.0889	&	0.052	&	&	0.5579	&	0.0934	&	0.1099	&	0.076	&	&	0.364	&	0.0835	&	0.1596	&	0.381	&	&	0.3887	&	0.084	&	0.1395	&	0.295	\\
& {\sf CF0}&		0.5067	&	0.0887	&	0.0889	&	0.052	&	&	0.5129	&	0.0987	&	0.0995	&	0.04	&	&	0.5029	&	0.1025	&	0.1025	&	0.048	&	&	0.5063	&	0.0908	&	0.0911	&	0.05	\\
& {\sf MW}1&		0.5088	&	0.0913	&	0.0917	&	0.005	&	&	0.5148	&	0.0993	&	0.1004	&	0.026	&	&	0.507	&	0.1074	&	0.1076	&	0.038	&	&	0.5112	&	0.0941	&	0.0947	&	0.033	\\
& {\sf MW}2&		0.5086	&	0.0905	&	0.0909	&	0.003	&	&	0.513	&	0.0994	&	0.1003	&	0.021	&	&	$\infty$	&	$\infty$	&	$\infty$	&	0.025	&	&	0.6415	&	0.1804	&	0.2293	&	0.087	\\
& {\sf DONG}&		$\infty$	&	$\infty$	&	$\infty$	&	0	&	&	$\infty$	&	$\infty$	&	$\infty$	&	0	&	&	0.5078	&	0.107	&	0.1073	&	0.037	&	&	0.603	&	0.1187	&	0.1572	&	0.09	\\
\cmidrule(r){2-21}
\multirow{5}{*}{\rotatebox[origin=c]{90}{$\alpha_1 = 1.00$}} 	& {\sf ML}&	0.7606	&	0.1132	&	0.2648	&	0.599	&	&	0.7121	&	0.1126	&	0.3092	&	0.735	&	&	1.1318	&	0.1377	&	0.1906	&	0.153	&	&	1.1226	&	0.1332	&	0.181	&	0.17	\\
& {\sf CF0}&	0.7606	&	0.1132	&	0.2648	&	0.599	&	&	1.002	&	0.2571	&	0.2571	&	0.051	&	&	1.0259	&	0.1331	&	0.1356	&	0.052	&	&	1.0204	&	0.1248	&	0.1264	&	0.052	\\
& {\sf MW}1&	0.6912	&	1.0619	&	1.1059	&	0.132	&	&	0.9878	&	0.2556	&	0.2559	&	0.038	&	&	1.0261	&	0.1363	&	0.1388	&	0.041	&	&	1.0209	&	0.134	&	0.1356	&	0.041	\\
& {\sf MW}2&	0.6519	&	1.1918	&	1.2416	&	0.074	&	&	0.9971	&	0.2562	&	0.2563	&	0.040	&	&	$\infty$ &	$\infty$	&	$\infty$	&	0.034	&	&	1.0952	&	0.1515	&	0.179	&	0.069	\\
& {\sf DONG}&	-$\infty$	&	$\infty$	&	$\infty$	&	0	&	&	-$\infty$	&	$\infty$	&	$\infty$	&	0	&	&	1.0257	&	0.1363	&	0.1387	&	0.039	&	&	0.9979	&	0.13	&	0.13	&	0.045	\\
 \cmidrule(r){2-21}
\multirow{8}{*}{\rotatebox[origin=c]{90}{$\beta = 1.00$}} 	& {\sf ML}&		1.5368	&	0.1392	&	0.5545	&	0.997	&	&	1.6078	&	0.1636	&	0.6294	&	0.995	&	&	1.4116	&	0.1296	&	0.4315	&	0.966	&	&	1.42	&	0.1388	&	0.4423	&	0.942	\\
& {\sf CF0}&		1.5368	&	0.1392	&	0.5545	&	0.997	&	&	1.0556	&	0.4665	&	0.4698	&	0.063	&	&	1.0361	&	0.1965	&	0.1998	&	0.047	&	&	1.0247	&	0.1754	&	0.1772	&	0.046	\\
& {\sf MW}1&		1.6867	&	2.1088	&	2.2178	&	0.133	&	&	1.0748	&	0.4728	&	0.4786	&	0.042	&	&	1.0287	&	0.2085	&	0.2104	&	0.052	&	&	1.0158	&	0.1913	&	0.1919	&	0.044	\\
& {\sf MW}2&		1.7579	&	2.3516	&	2.4707	&	0.071	&	&	1.0576	&	0.4756	&	0.4791	&	0.048	&	&	-$\infty$	&	$\infty$	&	$\infty$	&	0.031	&	&	0.7247	&	0.4179	&	0.5004	&	0.113	\\
& {\sf DONG}&		$\infty$	& $\infty$	&$\infty$	&	0	&	&	$\infty$	&	$\infty$	&	$\infty$	&	0	&	&	1.0261	&	0.208	&	0.2096	&	0.049	&	&	0.9828	&	0.2018	&	0.2026	&	0.049	\\
\cmidrule(r){3-21}
& {\it np}{\sf MW}1&		2.4255	&	2.3289	&	2.7305	&	0.196	&	&	1.2232	&	1.0193	&	1.0434	&	0.049	&	&	1.0064	&	0.1949	&	0.195	&	0.019	&	&	0.9701	&	0.1956	&	0.1978	&	0.023	\\
& {\it np}{\sf MW}2&		2.4552	&	2.4117	&	2.8168	&	0.083	&	&	1.1961	&	1.0294	&	1.0479	&	0.065	&	&	0.442	&	1.4108	&	1.5171	&	0.045	&	&	0.3403	&	0.5813	&	0.8793	&	0.148	\\
& {\it np}{\sf DONG}&		1.9991	&	2.0399	&	2.2714	&	0.141	&	&	2.0167	&	2.0646	&	2.3014	&	0.132	&	&	1.0033	&	0.1964	&	0.1965	&	0.02	&	&	0.9801	&	0.2076	&	0.2085	&	0.022	\\
 \cmidrule(r){2-21}
\multirow{8}{*}{\rotatebox[origin=c]{90}{$\rho = 0.50$}} 	& {\sf ML}&		{\it na}	&	{\it na}	&	{\it na}	&	{\it na}	&	&	{\it na}	&	{\it na}	&	{\it na}	&	{\it na}	&	&	{\it na}	&	{\it na}	&	{\it na}	&	{\it na}	&	&	{\it na}	&	{\it na}	&	{\it na}	&	{\it na}	\\	
& {\sf CF0}&	{\it na}	&	{\it na}	&	{\it na}	&	{\it na}	&	&	0.4929	&	0.3904	&	0.3905	&	0.06	&	&	0.4917	&	0.1973	&	0.1974	&	0.039	&	&	0.5022	&	0.1532	&	0.1533	&	0.052	\\	
& {\sf MW}1&		-0.1414	&	2.1096	&	2.205	&	0.131	&	&	0.4792	&	0.4002	&	0.4007	&	0.041	&	&	0.4979	&	0.2111	&	0.2112	&	0.038	&	&	0.5071	&	0.1662	&	0.1664	&	0.048	\\	
& {\sf MW}2&		-0.2119	&	2.3547	&	2.4599	&	0.073	&	&	0.4945	&	0.4031	&	0.4031	&	0.046	&	&	$\infty$	&	$\infty$	&	$\infty$	&	0.028	&	&	0.7162	&	0.4391	&	0.4894	&	0.069	\\	
& {\sf DONG}&		{\it na}	&	{\it na}	&	{\it na}	&	{\it na}	&	&	{\it na}	&	{\it na}	&	{\it na}	&	{\it na}	&	&	0.5017	&	0.2106	&	0.2106	&	0.037	&	&	0.6727	&	0.2327	&	0.2898	&	0.081	\\	
\cmidrule(r){3-21}
& {\it np}{\sf MW}1&		-0.186	&	1.1752	&	1.3608	&	0.154	&	&	0.4425	&	0.3942	&	0.3984	&	0.038	&	&	0.5182	&	0.2865	&	0.2871	&	0.022	&	&	0.5652	&	0.2385	&	0.2472	&	0.009	\\	
& {\it np}{\sf MW}2&		-0.2008	&	1.2147	&	1.4023	&	0.078	&	&	0.4528	&	0.3952	&	0.398	&	0.044	&	&	0.9953	&	1.6786	&	1.7501	&	0.034	&	&	1.036	&	0.7878	&	0.9529	&	0.025	\\	
& {\it np}{\sf DONG}&		0.0332	&	1.0067	&	1.1096	&	0.111	&	&	0.1543	&	0.9632	&	1.0233	&	0.064	&	&	0.5264	&	0.2882	&	0.2894	&	0.019	&	&	0.7514	&	0.3422	&	0.4247	&	0.029	\\	
 \cmidrule(r){2-21}
 \multirow{8}{*}{\rotatebox[origin=c]{90}{ASF $= 0.67/0.81$}} 	& {\sf ML}&		0.6934	&	0.039	&	0.0439	&	0.101	&	&	0.711	&	0.0396	&	0.0548	&	0.208	&	&	0.8558	&	0.0285	&	0.0502	&	0.407	&	&	0.8619	&	0.0292	&	0.0557	&	0.496	\\
& {\sf CF0}&		0.6934	&	0.039	&	0.0439	&	0.101	&	&	0.6732	&	0.0472	&	0.0472	&	0.089	&	&	0.8169	&	0.0347	&	0.0348	&	0.052	&	&	0.8162	&	0.0341	&	0.0341	&	0.047	\\
& {\sf MW}1&		0.6225	&	0.0563	&	0.0758	&	0.256	&	&	0.6746	&	0.0476	&	0.0477	&	0.061	&	&	0.8162	&	0.0356	&	0.0357	&	0.052	&	&	0.8156	&	0.0357	&	0.0357	&	0.048	\\
& {\sf MW}2&		0.6225	&	0.0564	&	0.0758	&	0.202	&	&	0.674	&	0.0476	&	0.0476	&	0.067	&	&	0.8210	&	0.0384	&	0.0390	&	0.007	&	&	0.8122	&	0.0385	&	0.0385	&	0.022	\\
& {\sf DONG}&		{\it na}	&	{\it na}	&	{\it na}	&	{\it na}	&	&	{\it na}	&	{\it na}	&	{\it na}	&	{\it na}	&	&	0.8157	&	0.0356	&	0.0357	&	0.049	&	&	0.8153	&	0.0352	&	0.0352	&	0.052	\\
\cmidrule(r){3-21}
& {\it np}{\sf MW}1&		0.6405	&	0.0517	&	0.0612	&	0.161	&	&	0.6615	&	0.0507	&	0.052	&	0.049	&	&	0.8052	&	0.0386	&	0.0396	&	0.045	&	&	0.796	&	0.0389	&	0.0431	&	0.054	\\
& {\it np}{\sf MW}2&		0.6375	&	0.052	&	0.063	&	0.179	&	&	0.66	&	0.0508	&	0.0525	&	0.073	&	&	0.721	&	0.0749	&	0.1197	&	0.456	&	&	0.7452	&	0.0694	&	0.0981	&	0.29	\\
& {\it np}{\sf DONG}&		0.6513	&	0.0512	&	0.0557	&	0.111	&	&	0.6551	&	0.0765	&	0.0786	&	0.125	&	&	0.8043	&	0.0383	&	0.0397	&	0.048	&	&	0.7997	&	0.0391	&	0.0418	&	0.048	\\
\hline\\[-13pt]
\bottomrule
\multicolumn{21}{p{21.5cm}}{\footnotesize {\textsf{Note}:} `$\infty$' means $>100$; while ASF = $a/b$ means ASF $= a$ for $\pi(z)=z$ and ASF $=b$ for $\pi(z)=z^2$;  {\sf ML} and {\sf CF}0 denote the ML estimator and the ML estimator with infeasible control function, respectively; {\sf MW}1/2 denote the new estimator with nonparametric/OLS first stage;  {\it np}{\sf MW}1/2 and {\it np}{\sf DONG} estimate the link function nonparametrically using the normalization $\alpha_1 = 1$ the other estimators use a probit link function with normalization $\sigma = 1$.}
\end{tabular}
\end{adjustbox}
\end{table}

\setlength{\tabcolsep}{1.4pt}
\begin{table} 
\caption{Monte Carlo results for $n=500$/$\rho = 0.00$}
\label{tab:mc2}
\vspace*{-.25cm}
\begin{adjustbox}{max width=\textwidth}
\begin{tabular}{lrccccccccccccccccccc} \toprule \\[-.5cm]  \hline\\[-.2cm]
&&\multicolumn{9}{c}{$\pi(z) = z$}&&\multicolumn{9}{c}{$\pi(z) =z^2$}\\[-.1cm]
\cmidrule(r){3-11} \cmidrule(r){13-21} \\[-.5cm]
&&\multicolumn{4}{c}{$V\sim\Phi$} & &  \multicolumn{4}{c}{$V\sim{\sf Gamma}(2,2)$} &&\multicolumn{4}{c}{$V\sim\Phi$} & &  \multicolumn{4}{c}{$V\sim{\sf Gamma}(2,2)$} \\[-.1cm]
 \cmidrule(r){3-7} \cmidrule(r){8-11} \cmidrule(r){13-17} \cmidrule(r){18-21} \\[-.5cm]
&  & \sf mean  &   \sf std   &  \sf rmse  & \sf size  &&  \sf mean  &   \sf std   &  \sf rmse  & \sf size & & \sf mean  &   \sf std   &  \sf rmse  & \sf size  &&  \sf mean  &   \sf std   &  \sf rmse  & \sf size\\  \cmidrule(r){2-21}
\multirow{5}{*}{\rotatebox[origin=c]{90}{$\alpha_0 = 0.50$}} 	& {\sf ML}&		0.5072	&	0.0855	&	0.0858	&	0.048	&	&	0.5083	&	0.0868	&	0.0871	&	0.048	&	&	0.5052	&	0.0789	&	0.0791	&	0.043	&	&	0.5047	&	0.0797	&	0.0799	&	0.059	\\
& {\sf CF0}&		0.5072	&	0.0855	&	0.0858	&	0.048	&	&	0.5127	&	0.0895	&	0.0904	&	0.05	&	&	0.505	&	0.0923	&	0.0924	&	0.05	&	&	0.5046	&	0.0851	&	0.0853	&	0.062	\\
& {\sf MW}1&		0.509	&	0.088	&	0.0885	&	0.01	&	&	0.5128	&	0.0895	&	0.0904	&	0.039	&	&	0.5053	&	0.0929	&	0.0931	&	0.035	&	&	0.5048	&	0.0861	&	0.0862	&	0.051	\\
& {\sf MW}2&		0.5094	&	0.0866	&	0.0871	&	0.002	&	&	0.5128	&	0.0896	&	0.0905	&	0.038	&	&	0.3754	&	0.4921	&	0.5076	&	0.031	&	&	0.4965	&	0.1476	&	0.1477	&	0.044	\\
& {\sf DONG}&		$\infty$	&	$\infty$	&	$\infty$	&	0	&	&	$\infty$	&	$\infty$	&	$\infty$	&	0	&	&	0.5053	&	0.0935	&	0.0936	&	0.038	&	&	0.5054	&	0.0961	&	0.0963	&	0.045	\\
\cmidrule(r){2-21}
\multirow{5}{*}{\rotatebox[origin=c]{90}{$\alpha_1 = 1.00$}} 	& {\sf ML}&		1.0154	&	0.115	&	0.116	&	0.055	&	&	1.0119	&	0.1177	&	0.1183	&	0.047	&	&	1.0157	&	0.1164	&	0.1174	&	0.046	&	&	1.0146	&	0.111	&	0.1119	&	0.053	\\
& {\sf CF0}&		1.0154	&	0.115	&	0.116	&	0.055	&	&	0.9925	&	0.2072	&	0.2074	&	0.055	&	&	1.0246	&	0.136	&	0.1382	&	0.055	&	&	1.0226	&	0.1249	&	0.127	&	0.053	\\
& {\sf MW}1&		0.9563	&	0.8831	&	0.8842	&	0.099	&	&	0.9923	&	0.2061	&	0.2063	&	0.04	&	&	1.025	&	0.1368	&	0.1391	&	0.048	&	&	1.0225	&	0.1257	&	0.1277	&	0.048	\\
& {\sf MW}2&		0.9432	&	0.9861	&	0.9877	&	0.059	&	&	0.9916	&	0.2078	&	0.2079	&	0.042	&	&	1.0264	&	0.1461	&	0.1484	&	0.01	&	&	1.0194	&	0.1158	&	0.1175	&	0.036	\\
& {\sf DONG}&		$\infty$	& $\infty$ &	$\infty$	&	0	&	&	-$\infty$	&	$\infty$	&	$\infty$	&	0	&	&	1.025	&	0.137	&	0.1392	&	0.049	&	&	1.0227	&	0.1286	&	0.1306	&	0.049	\\
 \cmidrule(r){2-21}
 \multirow{8}{*}{\rotatebox[origin=c]{90}{$\beta = 1.00$}} 	& {\sf ML}&		1.0179	&	0.1097	&	0.1112	&	0.056	&	&	1.0204	&	0.1195	&	0.1212	&	0.053	&	&	1.0166	&	0.0987	&	0.1	&	0.056	&	&	1.0178	&	0.1064	&	0.1079	&	0.05	\\
& {\sf CF0}&		1.0179	&	0.1097	&	0.1112	&	0.056	&	&	1.0684	&	0.3679	&	0.3742	&	0.053	&	&	1.033	&	0.1984	&	0.2011	&	0.059	&	&	1.032	&	0.1739	&	0.1768	&	0.057	\\
& {\sf MW}1&		1.1498	&	1.7485	&	1.7549	&	0.101	&	&	1.069	&	0.3701	&	0.3765	&	0.041	&	&	1.0324	&	0.2003	&	0.2029	&	0.051	&	&	1.0312	&	0.1764	&	0.1792	&	0.043	\\
& {\sf MW}2&		1.1768	&	1.9529	&	1.9609	&	0.069	&	&	1.0695	&	0.3754	&	0.3817	&	0.042	&	&	1.3082	&	1.0617	&	1.1056	&	0.038	&	&	1.0469	&	0.335	&	0.3383	&	0.037	\\
& {\sf DONG}&		-$\infty$	&	$\infty$	&	$\infty$	&	0	&	&	$\infty$	&	$\infty$	&	$\infty$	&	0	&	&	1.0323	&	0.2016	&	0.2042	&	0.052	&	&	1.0309	&	0.194	&	0.1965	&	0.041	\\
\cmidrule(r){3-21}
& {\it np}{\sf MW}1&		1.9463	&	2.2715	&	2.4607	&	0.14	&	&	1.1636	&	0.7732	&	0.7903	&	0.036	&	&	1.0058	&	0.1866	&	0.1867	&	0.025	&	&	0.9711	&	0.1764	&	0.1788	&	0.02	\\
& {\it np}{\sf MW}2&		1.9719	&	2.3492	&	2.5423	&	0.047	&	&	1.162	&	0.7814	&	0.798	&	0.037	&	&	1.0522	&	1.2408	&	1.2419	&	0.024	&	&	0.825	&	0.5075	&	0.5368	&	0.019	\\
&  {\it np}{\sf DONG}&		1.0066	&	1.6891	&	1.6891	&	0.064	&	&	0.9451	&	1.7005	&	1.7013	&	0.063	&	&	1.0034	&	0.188	&	0.1881	&	0.019	&	&	0.9921	&	0.19	&	0.1901	&	0.013	\\
\cmidrule(r){2-21}
\multirow{8}{*}{\rotatebox[origin=c]{90}{$\rho = 0.00$}} 	& {\sf ML}&		{\it na}	&	{\it na}	&	{\it na}	&	{\it na}	&	&	{\it na}	&	{\it na}	&	{\it na}	&	{\it na}	&	&	{\it na}	&	{\it na}	&	{\it na}	&	{\it na}	&	&	{\it na}	&	{\it na}	&	{\it na}	&	{\it na}	\\
& {\sf CF0}&		{\it na}	&	{\it na}	&	{\it na}	&	{\it na}	&	&	-0.0379	&	0.3103	&	0.3126	&	0.048	&	&	-0.0138	&	0.1989	&	0.1994	&	0.053	&	&	-0.0117	&	0.1458	&	0.1463	&	0.046	\\
& {\sf MW}1&		-0.1272	&	1.7551	&	1.7597	&	0.101	&	&	-0.0389	&	0.3154	&	0.3178	&	0.044	&	&	-0.0133	&	0.202	&	0.2024	&	0.043	&	&	-0.0112	&	0.1518	&	0.1523	&	0.044	\\
& {\sf MW}2&		-0.1534	&	1.9645	&	1.9705	&	0.062	&	&	-0.0391	&	0.319	&	0.3214	&	0.039	&	&	-0.3261	&	1.2118	&	1.255	&	0.037	&	&	-0.0231	&	0.3199	&	0.3208	&	0.036	\\
& {\sf DONG}&		{\it na}	&	{\it na}	&	{\it na}	&	{\it na}	&	&	{\it na}	&	{\it na}	&	{\it na}	&	{\it na}	&	&	-0.0132	&	0.2034	&	0.2038	&	0.043	&	&	-0.0107	&	0.209	&	0.2093	&	0.038	\\
\cmidrule(r){3-21}
&  {\it np}{\sf MW}1&		-0.6319	&	1.5354	&	1.6603	&	0.152	&	&	-0.0787	&	0.429	&	0.4362	&	0.029	&	&	0.02	&	0.2566	&	0.2573	&	0.015	&	&	0.0681	&	0.2094	&	0.2202	&	0.016	\\
&  {\it np}{\sf MW}2&		-0.649	&	1.5866	&	1.7142	&	0.056	&	&	-0.0771	&	0.4316	&	0.4384	&	0.038	&	&	-0.0145	&	1.4521	&	1.4522	&	0.021	&	&	0.2367	&	0.6096	&	0.6539	&	0.006	\\
&  {\it np}{\sf DONG}&		0.0006	&	1.1322	&	1.1322	&	0.054	&	&	0.0477	&	1.1341	&	1.1351	&	0.058	&	&	0.0227	&	0.2594	&	0.2604	&	0.018	&	&	0.0342	&	0.2735	&	0.2756	&	0.015	\\
 \cmidrule(r){2-21}
\multirow{8}{*}{\rotatebox[origin=c]{90}{ASF $0.69/0.84$}} 	& {\sf ML}&	0.6937	&	0.0382	&	0.0382	&	0.051	&	&	0.6941	&	0.038	&	0.038	&	0.054	&	&	0.8434	&	0.0263	&	0.0264	&	0.052	&	&	0.8432	&	0.0276	&	0.0276	&	0.058	\\
& {\sf CF0}&		{\it na}	&	{\it na}	&	{\it na}	&	{\it na}	&	&	0.6882	&	0.0379	&	0.0381	&	0.044	&	&	0.8406	&	0.0274	&	0.0274	&	0.043	&	&	0.8422	&	0.0287	&	0.0287	&	0.059	\\
& {\sf MW}1&		0.6313	&	0.0534	&	0.0808	&	0.234	&	&	0.688	&	0.038	&	0.0381	&	0.036	&	&	0.8405	&	0.0275	&	0.0275	&	0.032	&	&	0.842	&	0.0288	&	0.0288	&	0.048	\\
& {\sf MW}2&		0.6314	&	0.0534	&	0.0807	&	0.206	&	&	0.688	&	0.0379	&	0.0381	&	0.038	&	&	0.8422	&	0.0278	&	0.0278	&	0.001	&	&	0.8419	&	0.0287	&	0.0287	&	0.03	\\
& {\sf DONG}&		{\it na}	&	{\it na}	&	{\it na}	&	{\it na}	&	&	{\it na}	&	{\it na}	&	{\it na}	&	{\it na}	&	&	0.8404	&	0.0274	&	0.0274	&	0.034	&	&	0.8399	&	0.0282	&	0.0283	&	0.047	\\
\cmidrule(r){3-21}
&  {\it np}{\sf MW}1&		0.6358	&	0.0483	&	0.074	&	0.299	&	&	0.6777	&	0.0458	&	0.0479	&	0.052	&	&	0.8266	&	0.036	&	0.0389	&	0.042	&	&	0.8213	&	0.0412	&	0.0458	&	0.058	\\
&  {\it np}{\sf MW}2&		0.6325	&	0.0476	&	0.0761	&	0.359	&	&	0.6777	&	0.0457	&	0.0478	&	0.056	&	&	0.7654	&	0.0589	&	0.096	&	0.422	&	&	0.8015	&	0.0488	&	0.063	&	0.152	\\
&  {\it np}{\sf DONG}&		0.6517	&	0.0547	&	0.0679	&	0.183	&	&	0.643	&	0.0732	&	0.0881	&	0.193	&	&	0.8266	&	0.0361	&	0.0389	&	0.04	&	&	0.8223	&	0.0397	&	0.044	&	0.044	\\
\hline\\[-13pt]
\bottomrule
\multicolumn{21}{p{22cm}}{\footnotesize {\textsf{Note}:} `$\infty$' means $>100$; while ASF = $a/b$ means ASF $=a$ for $\pi(z)=z$ and ASF $=b$ for $\pi(z)=z^2$;  {\sf ML} and {\sf CF}0 denote the ML estimator and the ML estimator with infeasible control function, respectively; {\sf MW}1/2 denote the new estimator with nonparametric/OLS first stage;  {\it np}{\sf MW}1/2 and {\it np}{\sf DONG} estimate the link function nonparametrically using the normalization $\alpha_1 = 1$ the other estimators use a probit link function with normalization $\sigma = 1$.}
\end{tabular}
\end{adjustbox}
\end{table}

\setlength{\tabcolsep}{1.4pt}
\begin{table} 
\caption{Monte Carlo results for $n=$1,$000$/$\rho = 0.50$}
\label{tab:mc3}
\vspace*{-.25cm}
\begin{adjustbox}{max width=\textwidth}
\begin{tabular}{lrccccccccccccccccccc} \toprule \\[-.5cm]  \hline\\[-.2cm]
&&\multicolumn{9}{c}{$\pi(z) = z$}&&\multicolumn{9}{c}{$\pi(z) =z^2$}\\[-.1cm]
\cmidrule(r){3-11} \cmidrule(r){13-21} \\[-.5cm]
&&\multicolumn{4}{c}{$V\sim\Phi$} & &  \multicolumn{4}{c}{$V\sim{\sf Gamma}(2,2)$} &&\multicolumn{4}{c}{$V\sim\Phi$} & &  \multicolumn{4}{c}{$V\sim{\sf Gamma}(2,2)$} \\[-.1cm]
 \cmidrule(r){3-7} \cmidrule(r){8-11} \cmidrule(r){13-17} \cmidrule(r){18-21} \\[-.5cm]
&  & \sf mean  &   \sf std   &  \sf rmse  & \sf size  &&  \sf mean  &   \sf std   &  \sf rmse  & \sf size & & \sf mean  &   \sf std   &  \sf rmse  & \sf size  &&  \sf mean  &   \sf std   &  \sf rmse  & \sf size\\  \cmidrule(r){2-21}
\multirow{4}{*}{\rotatebox[origin=c]{90}{$\alpha_0 = 0.50$}} 	& {\sf ML}&		0.5057	&	0.0622	&	0.0624	&	0.047	&	&	0.5534	&	0.0656	&	0.0846	&	0.118	&	&	0.3637	&	0.0599	&	0.1489	&	0.634	&	&	0.3875	&	0.0583	&	0.1267	&	0.506	\\
& {\sf CF0}&		0.5057	&	0.0622	&	0.0624	&	0.047	&	&	0.5065	&	0.0676	&	0.0679	&	0.044	&	&	0.503	&	0.0713	&	0.0714	&	0.04	&	&	0.5044	&	0.0628	&	0.063	&	0.044	\\
& {\sf MW}1&		0.5063	&	0.0634	&	0.0637	&	0.01	&	&	0.5077	&	0.0679	&	0.0683	&	0.032	&	&	0.5044	&	0.0721	&	0.0722	&	0.031	&	&	0.5065	&	0.0638	&	0.0642	&	0.03	\\
& {\sf MW}2&		0.5063	&	0.0634	&	0.0637	&	0.003	&	&	0.5065	&	0.068	&	0.0683	&	0.031	&	&	$\infty$	&	$\infty$	&	$\infty$	&	0.123	&	&	0.6354	&	0.1246	&	0.184	&	0.152	\\
& {\sf DONG}&		$\infty$	&	$\infty$	&	$\infty$	&	0	&	&	-$\infty$	&	$\infty$	&	$\infty$	&	0	&	&	0.5049	&	0.0721	&	0.0722	&	0.033	&	&	0.5953	&	0.0806	&	0.1248	&	0.16	\\
\cmidrule(r){2-21}
\multirow{4}{*}{\rotatebox[origin=c]{90}{$\alpha_1 = 1.00$}} 	& {\sf ML}&		0.7544	&	0.0764	&	0.2572	&	0.88	&	&	0.7119	&	0.081	&	0.2993	&	0.945	&	&	1.1192	&	0.0979	&	0.1543	&	0.241	&	&	1.1111	&	0.0907	&	0.1434	&	0.245	\\
& {\sf CF0}&		0.7544	&	0.0764	&	0.2572	&	0.88	&	&	1.0083	&	0.1764	&	0.1766	&	0.061	&	&	1.0098	&	0.0941	&	0.0946	&	0.059	&	&	1.0059	&	0.0854	&	0.0856	&	0.053	\\
& {\sf MW}1&		0.7477	&	1.0677	&	1.0971	&	0.133	&	&	0.9999	&	0.1738	&	0.1738	&	0.043	&	&	1.011	&	0.0958	&	0.0964	&	0.059	&	&	1.0081	&	0.0909	&	0.0913	&	0.05	\\
& {\sf MW}2&		0.7412	&	1.1793	&	1.2074	&	0.083	&	&	1.0057	&	0.1757	&	0.1758	&	0.057	&	&	$\infty$	&	$\infty$	&	$\infty$	&	0.079	&	&	1.0769	&	0.1017	&	0.1275	&	0.099	\\
& {\sf DONG}&		$\infty$	&	$\infty$	&	$\infty$	&	0	&	&	-$\infty$&	$\infty$	&	$\infty$	&	0	&	&	1.0107	&	0.0959	&	0.0965	&	0.06	&	&	0.9857	&	0.0886	&	0.0897	&	0.045	\\
 \cmidrule(r){2-21}
\multirow{4}{*}{\rotatebox[origin=c]{90}{$\beta = 1.00$}} 	& {\sf ML}&		1.5139	&	0.1016	&	0.5238	&	1	&	&	1.5821	&	0.1174	&	0.5938	&	1	&	&	1.3974	&	0.0907	&	0.4076	&	0.999	&	&	1.4049	&	0.0938	&	0.4156	&	0.998	\\
& {\sf CF0}&		1.5139	&	0.1016	&	0.5238	&	1	&	&	1.0125	&	0.3139	&	0.3142	&	0.057	&	&	1.0199	&	0.1328	&	0.1343	&	0.051	&	&	1.014	&	0.119	&	0.1198	&	0.057	\\
& {\sf MW}1&		1.532	&	2.1164	&	2.1822	&	0.131	&	&	1.0256	&	0.3125	&	0.3135	&	0.043	&	&	1.0182	&	0.1372	&	0.1384	&	0.05	&	&	1.0128	&	0.1277	&	0.1284	&	0.046	\\
& {\sf MW}2&		1.5458	&	2.3367	&	2.3996	&	0.085	&	&	1.014	&	0.3155	&	0.3159	&	0.05	&	&	$\infty$	&	$\infty$	&	$\infty$	&	0.112	&	&	0.7201	&	0.2946	&	0.4064	&	0.17	\\
& {\sf DONG}&		$\infty$	&	$\infty$&	$\infty$	&	0	&	&	$\infty$	&	$\infty$	&	$\infty$	&	0	&	&	1.0169	&	0.1372	&	0.1383	&	0.047	&	&	0.9831	&	0.1337	&	0.1348	&	0.057	\\
\cmidrule(r){3-21}
&  {\it np}{\sf MW}1&		2.3366	&	2.3532	&	2.7063	&	0.189	&	&	1.0712	&	0.6029	&	0.6071	&	0.029	&	&	1.0072	&	0.1209	&	0.1211	&	0.015	&	&	0.9926	&	0.1202	&	0.1204	&	0.019	\\
&  {\it np}{\sf MW}2&		2.3073	&	2.4217	&	2.752	&	0.091	&	&	1.0574	&	0.617	&	0.6197	&	0.055	&	&	0.1954	&	1.0031	&	1.2859	&	0.074	&	&	0.5031	&	0.4351	&	0.6604	&	0.164	\\
&  {\it np}{\sf DONG}&		1.9066	&	2.0263	&	2.2199	&	0.129	&	&	2.0212	&	2.1051	&	2.3397	&	0.154	&	&	1.0077	&	0.1211	&	0.1213	&	0.014	&	&	1.0029	&	0.1238	&	0.1238	&	0.021	\\
 \cmidrule(r){2-21}
\multirow{8}{*}{\rotatebox[origin=c]{90}{$\rho = 0.50$}} 	& {\sf ML}&		{\it na}	&	{\it na}	&	{\it na}	&	{\it na}	&	&	{\it na}	&	{\it na}	&	{\it na}	&	{\it na}	&	&	{\it na}	&	{\it na}	&	{\it na}	&	{\it na}	&	&	{\it na}	&	{\it na}	&	{\it na}	&	{\it na}	\\
& {\sf CF0}&		{\it na}	&	{\it na}	&	{\it na}	&	{\it na}	&	&	0.5046	&	0.2691	&	0.2691	&	0.063	&	&	0.4945	&	0.1362	&	0.1363	&	0.051	&	&	0.4977	&	0.1096	&	0.1096	&	0.06	\\
& {\sf MW}1&		-0.0135	&	2.1225	&	2.1837	&	0.134	&	&	0.4945	&	0.2685	&	0.2686	&	0.038	&	&	0.4954	&	0.1403	&	0.1403	&	0.054	&	&	0.4967	&	0.1156	&	0.1156	&	0.051	\\
& {\sf MW}2&		-0.0268	&	2.3431	&	2.4016	&	0.085	&	&	0.5047	&	0.2724	&	0.2724	&	0.055	&	&	$\infty$	&	$\infty$	&	$\infty$	&	0.092	&	&	0.6974	&	0.3059	&	0.3641	&	0.091	\\
& {\sf DONG}&		{\it na}	&	{\it na}	&	{\it na}	&	{\it na}	&	&	{\it na}	&	{\it na}	&	{\it na}	&	{\it na}	&	&	0.4974	&	0.1399	&	0.1399	&	0.053	&	&	0.6562	&	0.1635	&	0.2261	&	0.161	\\
\cmidrule(r){3-21}
&  {\it np}{\sf MW}1&		-0.1597	&	1.1842	&	1.3556	&	0.168	&	&	0.4802	&	0.2416	&	0.2424	&	0.023	&	&	0.5075	&	0.1832	&	0.1834	&	0.037	&	&	0.5262	&	0.1553	&	0.1575	&	0.021	\\
&  {\it np}{\sf MW}2&		-0.146	&	1.2175	&	1.3783	&	0.09	&	&	0.4854	&	0.2451	&	0.2456	&	0.04	&	&	1.2471	&	1.1966	&	1.4106	&	0.031	&	&	0.8219	&	0.5483	&	0.6358	&	0.021	\\
&  {\it np}{\sf DONG}&		0.0588	&	1.0083	&	1.1006	&	0.11	&	&	0.128	&	0.9836	&	1.0516	&	0.082	&	&	0.5081	&	0.1838	&	0.184	&	0.032	&	&	0.7104	&	0.2174	&	0.3025	&	0.072	\\
 \cmidrule(r){2-21}
\multirow{8}{*}{\rotatebox[origin=c]{90}{ASD $= 0.67/0.81$}} 	& {\sf ML}&		0.6919	&	0.0272	&	0.0339	&	0.138	&	&	0.7084	&	0.0281	&	0.0463	&	0.358	&	&	0.8547	&	0.0206	&	0.0455	&	0.621	&	&	0.8608	&	0.0206	&	0.0511	&	0.723	\\
& {\sf CF0}&		{\it na}	&	{\it na}	&	{\it na}	&	{\it na}	&	&	0.6714	&	0.0336	&	0.0336	&	0.076	&	&	0.816	&	0.0248	&	0.0248	&	0.066	&	&	0.8157	&	0.0244	&	0.0244	&	0.065	\\
& {\sf MW}1&		0.6236	&	0.0534	&	0.0718	&	0.251	&	&	0.6724	&	0.0338	&	0.0338	&	0.062	&	&	0.816	&	0.0252	&	0.0252	&	0.053	&	&	0.8161	&	0.0255	&	0.0256	&	0.054	\\
& {\sf MW}2&		0.6169	&	0.054	&	0.0768	&	0.246	&	&	0.672	&	0.0338	&	0.0338	&	0.061	&	&	0.817	&	0.0391	&	0.0392	&	0.007	&	&	0.8118	&	0.0275	&	0.0276	&	0.016	\\
& {\sf DONG}&		{\it na}	&	{\it na}	&	{\it na}	&	{\it na}	&	&	{\it na}	&	{\it na}	&	{\it na}	&	{\it na}	&	&	0.8158	&	0.0252	&	0.0252	&	0.054	&	&	0.8165	&	0.0249	&	0.0251	&	0.049	\\
\cmidrule(r){3-21}
&  {\it np}{\sf MW}1&		0.6396	&	0.0482	&	0.0578	&	0.217	&	&	0.663	&	0.0372	&	0.0382	&	0.056	&	&	0.8078	&	0.0273	&	0.0281	&	0.046	&	&	0.8032	&	0.0283	&	0.0303	&	0.051	\\
&  {\it np}{\sf MW}2&		0.6363	&	0.0485	&	0.06	&	0.232	&	&	0.662	&	0.0373	&	0.0385	&	0.069	&	&	0.7169	&	0.0667	&	0.1178	&	0.514	&	&	0.766	&	0.0596	&	0.0766	&	0.213	\\
&  {\it np}{\sf DONG}&		0.6502	&	0.0464	&	0.0511	&	0.134	&	&	0.6532	&	0.0698	&	0.0722	&	0.143	&	&	0.8077	&	0.0274	&	0.0282	&	0.049	&	&	0.8071	&	0.0271	&	0.028	&	0.05	\\

\hline\\[-13pt]
\bottomrule
\multicolumn{21}{p{21.5cm}}{\footnotesize {\textsf{Note}:} `$\infty$' means $>100$; while ASF = $a/b$ means ASF $= a$ for $\pi(z)=z$ and ASF $=b$ for $\pi(z)=z^2$;  {\sf ML} and {\sf CF}0 denote the ML estimator and the ML estimator with infeasible control function, respectively; {\sf MW}1/2 denote the new estimator with nonparametric/OLS first stage;  {\it np}{\sf MW}1/2 and {\it np}{\sf DONG} estimate the link function nonparametrically using the normalization $\alpha_1 = 1$ the other estimators use a probit link function with normalization $\sigma = 1$.}
\end{tabular}
\end{adjustbox}
\end{table}

\setlength{\tabcolsep}{1.4pt}
\begin{table} 
\caption{Monte Carlo results for $n=$1,$000$/$\rho = 0.00$}
\label{tab:mc4}
\vspace*{-.25cm}
\begin{adjustbox}{max width=\textwidth}
\begin{tabular}{lrccccccccccccccccccc} \toprule \\[-.5cm]  \hline\\[-.2cm]
&&\multicolumn{9}{c}{$\pi(z) = z$}&&\multicolumn{9}{c}{$\pi(z) =z^2$}\\[-.1cm]
\cmidrule(r){3-11} \cmidrule(r){13-21} \\[-.5cm]
&&\multicolumn{4}{c}{$V\sim\Phi$} & &  \multicolumn{4}{c}{$V\sim{\sf Gamma}(2,2)$} &&\multicolumn{4}{c}{$V\sim\Phi$} & &  \multicolumn{4}{c}{$V\sim{\sf Gamma}(2,2)$} \\[-.1cm]
 \cmidrule(r){3-7} \cmidrule(r){8-11} \cmidrule(r){13-17} \cmidrule(r){18-21} \\[-.5cm]
&  & \sf mean  &   \sf std   &  \sf rmse  & \sf size  &&  \sf mean  &   \sf std   &  \sf rmse  & \sf size & & \sf mean  &   \sf std   &  \sf rmse  & \sf size  &&  \sf mean  &   \sf std   &  \sf rmse  & \sf size\\  \cmidrule(r){2-21}
\multirow{4}{*}{\rotatebox[origin=c]{90}{$\alpha_0 = 0.50$}} 	& {\sf ML}&		0.5057	&	0.0593	&	0.0596	&	0.051	&	&	0.5075	&	0.0599	&	0.0603	&	0.047	&	&	0.5067	&	0.057	&	0.0574	&	0.054	&	&	0.5066	&	0.0549	&	0.0553	&	0.045	\\
& {\sf CF0}&		0.5057	&	0.0593	&	0.0596	&	0.051	&	&	0.5087	&	0.061	&	0.0616	&	0.048	&	&	0.5045	&	0.0648	&	0.0649	&	0.052	&	&	0.5064	&	0.0579	&	0.0583	&	0.044	\\
& {\sf MW}1&		0.507	&	0.0602	&	0.0606	&	0.017	&	&	0.5084	&	0.061	&	0.0616	&	0.043	&	&	0.5046	&	0.065	&	0.0652	&	0.046	&	&	0.5063	&	0.058	&	0.0583	&	0.038	\\
& {\sf MW}2&		0.5065	&	0.0597	&	0.06	&	0.011	&	&	0.5085	&	0.061	&	0.0616	&	0.039	&	&	$\infty$	&	$\infty$	&	$\infty$	&	0.033	&	&	0.5057	&	0.1013	&	0.1015	&	0.049	\\
& {\sf DONG}&		$\infty$	&	$\infty$	&	$\infty$	&	0	&	&	$\infty$	&	$\infty$	&	$\infty$	&	0	&	&	0.5046	&	0.0651	&	0.0653	&	0.048	&	&	0.5057	&	0.0648	&	0.0651	&	0.035	\\
\cmidrule(r){2-21}
\multirow{4}{*}{\rotatebox[origin=c]{90}{$\alpha_1 = 1.00$}} 	& {\sf ML}&		1.0076	&	0.0802	&	0.0806	&	0.062	&	&	1.0093	&	0.0815	&	0.082	&	0.054	&	&	1.0068	&	0.0814	&	0.0817	&	0.053	&	&	1.007	&	0.0766	&	0.0769	&	0.051	\\
& {\sf CF0}&		1.0076	&	0.0802	&	0.0806	&	0.062	&	&	1.0078	&	0.1431	&	0.1433	&	0.056	&	&	1.0149	&	0.096	&	0.0971	&	0.056	&	&	1.0117	&	0.0873	&	0.0881	&	0.047	\\
& {\sf MW}1&		1.0181	&	0.8784	&	0.8786	&	0.107	&	&	1.0082	&	0.143	&	0.1433	&	0.053	&	&	1.0146	&	0.0959	&	0.097	&	0.057	&	&	1.0119	&	0.0877	&	0.0885	&	0.051	\\
& {\sf MW}2&		1.0188	&	0.9403	&	0.9405	&	0.063	&	&	1.0081	&	0.1442	&	0.1444	&	0.054	&	&	$\infty$	&	$\infty$	&	$\infty$	&	0.014	&	&	1.0078	&	0.0787	&	0.0791	&	0.041	\\
& {\sf DONG}&		-$\infty$	&	$\infty$	&	$\infty$	&	0	&	&	-$\infty$	&	$\infty$	&	$\infty$	&	0	&	&	1.0147	&	0.0959	&	0.0971	&	0.057	&	&	1.0128	&	0.0919	&	0.0928	&	0.052	\\
 \cmidrule(r){2-21}
\multirow{4}{*}{\rotatebox[origin=c]{90}{$\beta = 1.00$}} 	& {\sf ML}&		1.0089	&	0.0789	&	0.0794	&	0.063	&	&	1.0096	&	0.0847	&	0.0852	&	0.059	&	&	1.0081	&	0.0686	&	0.069	&	0.049	&	&	1.0099	&	0.0748	&	0.0754	&	0.056	\\
& {\sf CF0}&		1.0089	&	0.0789	&	0.0794	&	0.063	&	&	1.0167	&	0.2414	&	0.242	&	0.048	&	&	1.0242	&	0.139	&	0.1411	&	0.064	&	&	1.0183	&	0.1186	&	0.12	&	0.051	\\
& {\sf MW}1&		0.9887	&	1.737	&	1.7371	&	0.117	&	&	1.0153	&	0.2424	&	0.2429	&	0.043	&	&	1.0237	&	0.1388	&	0.1408	&	0.055	&	&	1.0185	&	0.1196	&	0.121	&	0.047	\\
& {\sf MW}2&		0.987	&	1.8678	&	1.8679	&	0.067	&	&	1.0153	&	0.2438	&	0.2443	&	0.045	&	&	-$\infty$	&	$\infty$	&	$\infty$	&	0.029	&	&	1.0155	&	0.2289	&	0.2294	&	0.044	\\
& {\sf DONG}&		$\infty$	&	$\infty$	&	$\infty$	&	0	&	&	$\infty$	&	$\infty$	&	$\infty$	&	0	&	&	1.0238	&	0.1392	&	0.1412	&	0.054	&	&	1.0206	&	0.1334	&	0.135	&	0.052	\\
 \cmidrule(r){3-21}
&  {\it np}{\sf MW}1&		1.6667	&	2.1902	&	2.2894	&	0.129	&	&	1.031	&	0.4543	&	0.4554	&	0.034	&	&	1.0075	&	0.1234	&	0.1236	&	0.022	&	&	0.9916	&0.1114	&0.1117	&0.014	\\
&  {\it np}{\sf MW}2&		1.743	&	2.2648	&	2.3836	&	0.053	&	&	1.0315	&	0.4577	&	0.4587	&	0.038	&	&	1.0214	&	0.9369	&	0.9371	&	0.023	&	&	0.9117	&0.333	&0.3445	&0.019	\\
&  {\it np}{\sf DONG}&		0.9083	&	1.6257	&	1.6283	&	0.06	&	&	0.9142	&	1.6254	&	1.6276	&	0.051	&	&	1.0069	&	0.1225	&	0.1227	&	0.023	&	&	1.007	&0.11	&0.1102	&0.018	\\
 \cmidrule(r){2-21}
\multirow{8}{*}{\rotatebox[origin=c]{90}{$\rho = 0.00$}} 	& {\sf ML}&		{\it na}	&	{\it na}	&	{\it na}	&	{\it na}	&	&	{\it na}	&	{\it na}	&	{\it na}	&	{\it na}	&	&	{\it na}	&	{\it na}	&	{\it na}	&	{\it na}	&	&	{\it na}	&	{\it na}	&	{\it na}	&	{\it na}	\\	
& {\sf CF0}&		{\it na}	&	{\it na}	&	{\it na}	&	{\it na}	&	&	-0.0042	&	0.2091	&	0.2091	&	0.04	&	&	-0.0161	&	0.142	&	0.1429	&	0.066	&	&	-0.0071	&	0.1031	&	0.1034	&	0.053	\\	
& {\sf MW}1&		0.0233	&	1.7397	&	1.7399	&	0.115	&	&	-0.0028	&	0.2104	&	0.2104	&	0.034	&	&	-0.0156	&	0.1416	&	0.1425	&	0.057	&	&	-0.0075	&	0.1041	&	0.1044	&	0.049	\\	
& {\sf MW}2&		0.0253	&	1.8696	&	1.8698	&	0.067	&	&	-0.0029	&	0.2121	&	0.2121	&	0.04	&	&	$\infty$	&	$\infty$	&	$\infty$	&	0.031	&	&	-0.0028	&	0.224	&	0.2241	&	0.044	\\	
& {\sf DONG}&		{\it na}	&	{\it na}	&	{\it na}	&	{\it na}	&	&	{\it na}	&	{\it na}	&	{\it na}	&	{\it na}	&	&	-0.0157	&	0.1422	&	0.143	&	0.058	&	&	-0.0106	&	0.1442	&	0.1446	&	0.049	\\	
 \cmidrule(r){3-21}
&  {\it np}{\sf MW}1&		-0.4386	&	1.4602	&	1.5247	&	0.12	&	&	-0.014	&	0.2593	&	0.2596	&	0.029	&	&	0.0028	&	0.17	&	0.17	&	0.024	&	&	0.0278	&0.1335	&0.1363	&0.017	\\	
&  {\it np}{\sf MW}2&		-0.489	&	1.5091	&	1.5864	&	0.059	&	&	-0.0142	&	0.2602	&	0.2606	&	0.033	&	&	0.0004	&	1.0906	&	1.0906	&	0.019	&	&	0.1163&	0.382&	0.3993&	0.014	\\	
&  {\it np}{\sf DONG}&		0.0668	&	1.0808	&	1.0829	&	0.056	&	&	0.057	&	1.0936	&	1.095	&	0.047	&	&	0.0042	&	0.168	&	0.168	&	0.023	&	&	0.004&	0.1667&	0.1668	&0.009\\	
 \cmidrule(r){2-21}
\multirow{8}{*}{\rotatebox[origin=c]{90}{ASF$= 0.69/0.84$}} 	& {\sf ML}&		0.6919	&	0.027	&	0.027	&	0.054	&	&	0.6926	&	0.0277	&	0.0278	&	0.054	&	&	0.8429	&	0.0195	&	0.0196	&	0.073	&	&	0.8431	&	0.0201	&	0.0202	&	0.059	\\
& {\sf CF0}&		{\it na}	&	{\it na}	&	{\it na}	&	{\it na}	&	&	0.6894	&	0.0276	&	0.0277	&	0.05	&	&	0.8418	&	0.0204	&	0.0204	&	0.067	&	&	0.8427	&	0.021	&	0.021	&	0.069	\\
& {\sf MW}1&		0.6318	&	0.0486	&	0.0761	&	0.244	&	&	0.6893	&	0.0277	&	0.0278	&	0.047	&	&	0.8418	&	0.0205	&	0.0205	&	0.061	&	&	0.8426	&	0.0211	&	0.0211	&	0.064	\\
& {\sf MW}2&		0.6268	&	0.0492	&	0.0803	&	0.228	&	&	0.6893	&	0.0277	&	0.0277	&	0.045	&	&	0.7950	&	0.0512	&	0.0688	&	0.057	&	&	0.8422	&	0.0207	&	0.0208	&	0.037	\\
& {\sf DONG}&		{\it na}	&	{\it na}	&	{\it na}	&	{\it na}	&	&	{\it na}	&	{\it na}	&	{\it na}	&	{\it na}	&	&	0.8418	&	0.0204	&	0.0205	&	0.06	&	&	0.8416	&	0.0205	&	0.0205	&	0.052	\\
 \cmidrule(r){3-21}
&  {\it np}{\sf MW}1&		0.6348	&	0.0425	&	0.0699	&	0.396	&	&	0.6806	&	0.035	&	0.0363	&	0.059	&	&	0.832	&	0.0286	&	0.0299	&	0.05	&	&	0.8301&	0.03	&0.0319	&0.038	\\
&  {\it np}{\sf MW}2&		0.6325	&	0.0433	&	0.0722	&	0.436	&	&	0.6807	&	0.0349	&	0.0362	&	0.06	&	&	0.7818	&	0.0504	&	0.0777	&	0.339	&	&	0.8183	&0.0361&	0.0426&	0.083	\\
& {\it np}{\sf DONG}&		0.6507	&	0.046	&	0.0607	&	0.231	&	&	0.6444	&	0.066	&	0.0804	&	0.212	&	&	0.8319	&	0.0286	&	0.0299	&	0.05	&	&	0.8312&	0.0284&0.0301&	0.034	\\
\hline\\[-13pt]
\bottomrule
\multicolumn{21}{p{22cm}}{\footnotesize {\textsf{Note}:} `$\infty$' means $>100$; while ASF = $a/b$ means ASF $= a$ for $\pi(z)=z$ and ASF $=b$ for $\pi(z)=z^2$;  {\sf ML} and {\sf CF}0 denote the ML estimator and the ML estimator with infeasible control function, respectively; {\sf MW}1/2 denote the new estimator with nonparametric/OLS first stage;  {\it np}{\sf MW}1/2 and {\it np}{\sf DONG} estimate the link function nonparametrically using the normalization $\alpha_1 = 1$ the other estimators use a probit link function with normalization $\sigma = 1$.}
\end{tabular}
\end{adjustbox}
\end{table}

In each of the 1,000 Monte Carlo repetitions, we take {\sf IID} draws $\{Y_i, Z_i, D_i\}_{i=1}^n$, where $n \in \{500,$ 1,000$\}$, based on the specifications discussed earlier for estimation and inference. We consider eight different estimators:

\begin{enumerate}
    \item[(1)] The (biased) probit ML estimator that neglects endogeneity ({\sf ML}).
   \item[(2)] The infeasible LIML probit estimator, which includes $m(V)$ as a control function ({\sf CF}0).
\item [] The feasible LIML probit estimator, where our rank-based estimate $\eta_{i,n}$ of $m(V)$ uses residuals $V_{i,n}$ based on:
   \begin{enumerate}
\item[(3)] a nonparametrically estimated first stage ({\sf MW}1), using the {\sf gam} function for additive models with default settings from the {\sf R} package {\sf mgcv}, or
\item[(4)] a linear first stage estimated via OLS ({\sf MW}2).
\end{enumerate}
   \item[(5)] The estimator from \cite{dong:2010} ({\sf DONG}), which uses the first-stage residual $V_{i,n}$, nonparametrically estimated as in {\sf MW}1, as a control function.
\end{enumerate}
Finally, we consider also the three last estimators (3)-(5) with the link function estimated using the Nadaraya-Watson estimator. These three additional estimators are denoted by (6) {\it np}{\sf MW}1, (7) {\it np}{\sf MW}2, and (8) {\it np}{\sf DONG}. The kernel-based estimation of the link function and the ASF (see Section 3.4.) is implemented using the {\sf R} function {\sf kreg} with default settings.

For estimators (1)-(5), we use the normalization $\sigma = 1$, while, following the literature, for (6)-(8), we use $\alpha_1 = 1$. Note that, due to the local level specification of the nonparametric estimator of the link function, estimators (6)-(8) do not include a constant.

We report the following metrics for each estimator: the mean, the standard deviation,
the root-mean-squared error, the empirical size of a two-sided $t$-test at the nominal significance level of 5\% for the estimators of $\theta$ and the ASF evaluated at the mean of $X$.
For estimators (3)-(5), we compute test statistics using bootstrap standard errors with 499 repetitions. For computational reasons, estimators (6)-(8) use 99 bootstrap repetitions.

Table \ref{tab:mc1} shows that in case of endogeneity ($\rho = 0.5$), the proposed endogeneity correction does its job as long as Assumption \ref{ass:ident} is satisfied. The naïve probit estimator ({\sf ML}) displays severe bias and size distortions unless endogeneity is absent (i.e.\ $\rho = 0.0$, see Table \ref{tab:mc2}), in which case it coincides with {\sf CF}0. As expected, the case of a linear reduced form in conjunction with $H=G$ (i.e.\ violation of Assumption \ref{ass:ident}) leads to non-identification due to collinearity. Evidently, in this case {\sf CF}0 is not defined as collinearity becomes perfect. We note that {\sf MW}2 has severe problems in case the nonlinear first-stage is misspecified, while the efficiency loss of {\sf MW}1 that estimates the first stage nonparametrically relative to its infeasible counterpart {\sf CF}0 seems acceptable. When the link function is estimated nonparametrically, estimation precision--though still satisfactory--is generally lower compared to the probit specifications. As shown in Tables \ref{tab:mc3} and \ref{tab:mc4}, both estimation performance and size control improve as the sample size increases from $n = 500$ to $n = $1,000. 

Finally, the estimator proposed by \cite{dong:2010}, which uses the nonparametrically estimated innovation $V$ as a control function, is more efficient than {\sf MW}1 in the setting where  $\pi(z) = z^2$ and $V \sim \Phi$, so that $m(V) = V$. This is because the rank-based estimation of $m(\cdot)$, as employed by {\sf MW}1, is superfluous here. In all other scenarios, in particular if the dependence in the first step is linear, {\sf MW}1 outperforms {\sf DONG}.

\section{Application to Insolvency Risk}\label{sec:emp}

We consider German administrative data from the \textit{Forschungsdatenzentren der statistischen Ämter des Bundes und der Länder}, which contains all German companies (\textit{Rechtseinheiten}) in the year 2018 and 2019. The general task is to model insolvency risk, which is a currently relevant topic, see e.g. \citet{weissbachwied:2022}. In particular, we are interested in measuring the influence of company growth on insolvency risk: Is strong growth an indicator for a healthy company or does strong growth imply substantial risk? Dependencies between company growth and insolvency risk have been of interest in corporate development for a long time, see \citet{bensoussan:1981}, \citet{santanna:2017} or \citet{xuezhouetal:2022}.

With this question, potential endogeneity issues arise: There might be reverse causality (if insolvency lies on the table, employees might leave the company) or the existence of a latent relevant variable which measures the current quality of the management and related aspects.

The dependent variable is the indicator variable if an insolvency case starts in 2019. While such insolvency cases can take several years, typically a five-digit number of companies actually becomes insolvent in Germany per year (\citealp{weissbachwied:2022}). The base model is given by
\begin{align*}
{\sf P}(InsolvencyCaseStarts&2019_i)  \\
 = \Phi(\beta_0 & + \beta_1 SalesGrowth2019_i    + \beta_2 EmployeeGrowth2019_i \\
 &  + \beta_3 Sales2018_i + \beta_4 Employees2018_i + Controls_i).
\end{align*}

Potential endogenous variables are the sales growth from 2018 to 2019 and the employee growth from 2018 to 2019. Exogenous controls are the sales in 2018, the employees in 2018 the German state and the legal status (\textit{Rechtsform}). Companies whose sales in 2018 are below and above the 5\% and 95\% quantiles of these sales are excluded. This leads to a sample size of $n=1$,$131$,$230$. In the sample, for 3,412 companies, an insolvency case starts in 2019.

We believe that it is reasonable to assume the existence of normally distributed latent terms, which determine the sales and employee growths. These terms refer to ``management intelligence'' and there is evidence for the fact that such intelligence-related terms are normally distributed (\citealp{bmw:24}). On the other hand, growth values are typically non-normally distributed, so that our nonlinearity condition should be fulfilled. In our dataset, there is no explicit information about such terms. Moreover, no instruments such as external or internal firm growth as considered in \citet{xuezhouetal:2022} are available. 

We show in Table \ref{table:app} the results for a probit regression without control terms ({\sf ML}), for control terms (one for both endogenous variables) with a nonparametric first step ({\sf MW}1) and a linear first step ({\sf MW}2). Moreover, we provide a comparison with the Probit estimator from \cite{dong:2010}. The estimates for the state and legal status are omitted for brevity. The standard errors for the regressions with control terms are obtained via bootstrap (99 replications), the other ones via the standard Fisher information from the likelihood.

\begin{table}[!h!]\footnotesize
\begin{center}
\caption{\small Estimation results}  \label{table:app}
\vspace*{-.25cm}
\begin{tabular}{rcccccccc} \toprule \\[-3ex]
 & \multicolumn{2}{c}{\sf ML} & \multicolumn{2}{c}{{\sf MW}1} & \multicolumn{2}{c}{{\sf MW}2} & \multicolumn{2}{c}{ \sf DONG} \\ 
 \cmidrule(l){2-3} \cmidrule(l){4-5} \cmidrule(l){6-7} \cmidrule(l){8-9}
            & \textit{estim} & \textit{t stat}  &   \textit{estim} & \textit{t stat} &      \textit{estim} & \textit{t stat} &  \textit{estim} & \textit{t stat} \\ 
             \cmidrule(l){2-9}
$\beta_1$ & -0.213 & -11.67 & 0.001 & 1.54   & 0.001 & 1.27 & 0.052 & 0.349  \\[-1ex]
        & {\scriptsize (0.018)} &  & {\scriptsize (0.001)}  &  & {\scriptsize (0.001)} & & {\scriptsize (0.148)} &  \\
         $\beta_2$ & -0.988 & -42.50  & 0.011 & 6.43 & 0.012 & 6.27 & -0.156 & -0.157 \\[-1ex]
         & {\scriptsize (0.023)} &  & {\scriptsize (0.002)}  &  & {\scriptsize (0.002)} & & {\scriptsize (0.994)} &  \\
 \cmidrule(l){2-3} \cmidrule(l){4-5} \cmidrule(l){6-7} \cmidrule(l){8-9}
  $\beta_0$ & -1.673 & -42.54 &  -3.045 & -83.25 & -3.017 & -78.88 & -2.897 & -2.968 \\[-1ex]
            & {\scriptsize (0.039)} &  & {\scriptsize (0.037)}  &  & {\scriptsize (0.038)} & & {\scriptsize (0.976)} &   \\
  $\beta_3$ & $>$ -0.001 & -2.84 & $<$ 0.001 & 14.89 & $<$ 0.001 & 14.77 & $<$ 0.001 & 2.922 \\[-1ex]
            & {\scriptsize ($<$ 0.001)} &  & {\scriptsize ($<$ 0.001)}  &  & {\scriptsize ($<$ 0.001)} & & {\scriptsize ($<$ 0.001)} &  \\
  $\beta_4$ & $<$ 0.001 & 1.69 & $>$ -0.001 & -0.55 & $<$ 0.001 & 0.10 & $>$ -0.001 & -0.221 \\[-1ex]
            & {\scriptsize ($<$ 0.001)} &  & {\scriptsize ($<$ 0.001)}  &  & {\scriptsize ($<$ 0.001)} & & {\scriptsize (0.003)} &  \\
 \cmidrule(l){2-3} \cmidrule(l){4-5} \cmidrule(l){6-7} \cmidrule(l){8-9}
  $\rho_1$ &   &   & -0.227 & -21.02 & -0.236 & -21.21 & -0.267 & -2.294 \\[-1ex]
            &  &  & {\scriptsize (0.011)}  &  & {\scriptsize (0.011)} & & {\scriptsize (0.116)} & \\
              $\rho_2$ &   &   & -0.312 & -35.78 & -0.300 & -33.93 & -0.841 & -0.852 \\[-1ex]
            &  &  & {\scriptsize (0.009)}  &  & {\scriptsize (0.009)} & & {\scriptsize (0.986)} & \\
\bottomrule 
\end{tabular}
\end{center}
\end{table}

The results indicate that the control terms are very relevant. Without including them, the estimates for $\beta_1$ and $\beta_2$ are statistically significantly negative. With the terms, they become positive in most cases, whereas the $t$-statistics decrease in absolute values. Notably, the estimates for the control terms are statistically significantly negative.

This supports the interpretation that the control terms capture the current quality of management and related factors. A higher management quality reduces the likelihood of insolvency proceedings. When this factor is accounted for, an increase in sales and employment raises the probability of insolvency, likely because firms take on greater risks in pursuit of growth.

Interestingly, the standard errors of the \citet{dong:2010} estimates are substantially larger than those from our approach. This suggests that the dependence structure in the first step of our model is more linear than nonlinear. This conclusion is further supported by the fact that our estimator's results remain consistent between the linear and nonparametric specifications in the first step. More evidence is provided by additional plots of the fitted first stage values, which are not reported.

Similar results were obtained in \citet{xuezhouetal:2022}. These authors consider a partly similar model, but use a different estimation approach. In the same spirit as our analysis, their estimates for firm growth increase, once ``mediation variables'' (with negative coefficient estimates) are included into the model.

\section{Summary and Outlook}
This paper addresses a gap in the literature by proposing a rank-based endogeneity correction for binary outcome models in the presence of endogeneity, without relying on external instruments. The approach allows for both linear and nonlinear dependence between endogenous and exogenous regressors. While we focus on the case of a known link function, we also discuss potential extensions, including methods for handling unknown link functions, drawing inspiration from \cite{klein:93}, \cite{blundell:2004}, and \cite{rothe2010nonparametric}. Another promising direction for future research is the extension to distribution regression models, which build on binary outcome models as discussed recently by \cite{wied:2024}.

\section{Declarations}

\subsection{Data Availability Statement }

We use a unique dataset from the Research Data Centres of the Federal Statistical Office and Statistical Offices of the Federal States of Germany, which is not publicly available. Fee-based access can be granted by signing a contract.

\subsection{Funding and Competing Interests}

All authors certify that they have no affiliations with or involvement in any organization or entity with any financial interest or non-financial interest in the subject matter or materials discussed in this manuscript. The authors have no funding to report.

   
    \renewcommand{\theequation}{A.\arabic{equation}}
    \renewcommand{\thesection}{A}
     \section{Proofs}
    \setcounter{equation}{0}
     \noindent {\bf Proof of Proposition \ref{prop:consistency}.} The claim follows if we can show that the objective function ${\cal L}_n(\theta)$ is uniformly close to the infeasible objective ${\cal L}_{0,n}(\theta)$ that uses the unknown control function $\eta_i$ in place of $\eta_{i,n}$. To that end, let $\bar \eta_{i,n}$ (random) be on the line segment connecting $\eta_{i,n}$ and $\eta_i$. Then, by the mean-value theorem and Cauchy-Schwarz, we get
\begin{align}
\ssup\limits_{\theta \in \Theta}|{\cal L}_n(\theta) & - {\cal L}_{0,n}(\theta)| \nonumber\\
\leq |\rho| \,& \left[\ssup\limits_{\theta \in \Theta}\frac{1}{n}\sum_{i=1}^{n} \psi^2(\theta;Y_i,X_i,\bar\eta_{i,n})\right]^{1/2} \left[\frac{1}{n}\sum_{i=1}^{n}(\eta_{i,n}-\eta_i)^2\right]^{1/2}.\label{eq:A1}
\end{align}
As argued in the treatment of their term `$B$' in the proof of \citet[Theorem 3.4]{zhao:2020} (which builds upon \citealp[Proposition F.7]{zhao:19}), we get $\sum_{i=1}^{n}(\tilde\eta_{i,n}-\eta_i)^2 = o_p(n)$. Hence, as by Assumption \ref{ass:ranks} $\sum_{i=1}^{n}(\tilde\eta_{i,n}-\eta_{i,n})^2 = o_p(n)$, it follows from the triangle inequality for the last term on the right-hand side of Eq. \eqref{eq:A1}, $\sum_{i=1}^{n}(\eta_{i,n}-\eta_i)^2 = o_p(n).$ Moreover, we obtain from Assumption  \ref{ass:psi}
\[
\ssup\limits_{\theta \in \Theta}\frac{1}{n}\sum_{i=1}^{n} \psi^2(\theta;Y_i,X_i,\bar\eta_{i,n}) \leq \ssup\limits_{\theta \in \Theta}\frac{1}{n}\sum_{i=1}^{n}\ssup_{t_i: |t_i-\eta_i|\leq b_n}\psi^2(\theta;Y_i,X_i,t_i)= O_p(1).
\]
This shows $\ssup\limits_{\theta \in \Theta}|{\cal L}_n(\theta) - {\cal L}_{0,n}(\theta)| = o_p(1)$ and the claim follows. \hfill$\square$\\

 \noindent{\bf Proof of Proposition \ref{prop:score}}. Define $A_1 \coloneqq n^{-1/2}\sum_{i=1}^ns_0(Y_i,X_i,\eta_i)$ and note that $A_1 \rightarrow_d {\cal A}
 _1 =_d \mathcal{N}(0,\Omega_1)$. Next, consider
\begin{equation}\nonumber
\begin{split}
\frac1{\sqrt{n}} \sum_{i=1}^n [s_0(Y_i,X_i,\eta_{i,n}) & -  s_{0}(Y_i,X_i,\eta_i)] \\
= \,&  \frac1{\sqrt{n}} \sum_{i=1}^n\begin{bmatrix} X_i \\ \eta_i \end{bmatrix} [\psi_0(Y_i,X_i,\eta_{i,n})-\psi_0(Y_i,X_i,\eta_i)]  \\
\,& +  \frac1{\sqrt{n}} \sum_{i=1}^n\begin{bmatrix} 0_m \\ \eta_{i,n}-\eta_{i} \end{bmatrix} [\psi_0(Y_i,X_i,\eta_{i,n})-\psi_0(Y_i,X_i,\eta_i)] \\
\,& +  \frac1{\sqrt{n}} \sum_{i=1}^n\begin{bmatrix} 0_m \\ \eta_{i,n}- \eta_i \end{bmatrix} \psi_0(Y_i,X_i,\eta_{i}) \\
\eqqcolon \,& A+B+C,
 \end{split} 
\end{equation}
say. Begin with $A$ and note that, by a first-order Taylor expansion, we obtain
\begin{equation}\nonumber
\begin{split}
A = \,& \frac\rho{\sqrt n}\sum_{i=1}^n S_0(Y_i,X_i,\eta_i)(\tilde\eta_{i,n}-\eta_{i})\\
\,& + \frac\rho{\sqrt{n}} \sum_{i=1}^nS_0(Y_i,X_i,\eta_i) (\eta_{i,n}-\tilde\eta_{i,n}) + o_p(1) \eqqcolon  \rho(A_2+A_3) + o_p(1),
 \end{split} 
\end{equation}
with $A_2$ and $A_3$ being implicitly defined. Begin with $A_2$ and note that $\eta_i = \Phi^{-1}(U_i)$, with $U_i = G(V_i)$ being an {\sf IID} sequence of ${\sf Unif}[0,1]$ variates. Let $K_n$ denote the empirical cdf of $U_i$. Then, there exists an ordering of the indices
$\{1,\dots ,n\}$, such that, by a first-order Taylor expansion, we obtain for $-A_2$
\begin{equation}\nonumber
\begin{split}
 \frac1{\sqrt n}&\sum_{i=1}^n S_0(Y_i,X_i,\Phi^{-1}(K_n^{-1}(i/n)))(\Phi^{-1}(K_n^{-1}(i/n))-\Phi^{-1}(i/(n+1))) \\
= \,& \frac1{\sqrt n}\sum_{i=1}^n S_0(Y_i,X_i,\Phi^{-1}(K_n^{-1}(i/n)))\frac{K_n^{-1}(i/(n+1))-i/(n+1)}{\phi(\Phi^{-1}(i/(n+1)))}+o_p(1) \\
\,&\ \rightarrow_d \int_0^1 \frac{\Ex[S_0(Y,X,\Phi^{-1}(U) \mid U = u] B(u)}{\phi(\Phi^{-1}(u))} {\sf d}u \eqqcolon {\cal A}_2,
 \end{split} 
\end{equation}
almost surely, for a standard Brownian Bridge $B(\cdot)$ so that $\var[{\cal A}_2] = \Omega_2$. Here we used a similar argument to the proof of \citet[Proposition 3.1]{bmw:24}. Finally, $A_3 \rightarrow_d {\cal A}_3$, ${\cal A}_3 \sim \mathcal{N}(0,\Omega_3)$ follows by Assumptions \ref{ass:ranks} and \ref{ass:linrep}. The claim thus follows because, by Cauchy-Schwarz and the previous result, $B$ and $C$ are $o_p(1)$ and $\llim\cov[A_j,A_i] = 0$, $i\neq j$.  \hfill$\square$\\

\noindent{\bf Proof of Proposition \ref{prop:normal}}. Since $\theta_n$ is a solution of the maximisation problem ${\cal L}_n(\theta)$ it follows that $\sum_{i=1}^ns(\theta_n;Y_i,X_i,\eta_{i,n})=0_{k+2}$ and, by a first order Taylor expansion about $\theta_0$, we get
$$\sqrt{n}(\theta_n-\theta_0) = \left[-\frac1{n}\sum_{i=1}^nH(\theta_0;Y_i,X_i,\eta_{i})+O_p(\vert \theta_0-\theta_n\vert^2)\right]^{-1}\frac1{\sqrt n}\sum_{i=1}^ns(\theta_0;Y_i,X_i, \eta_{i,n}),$$
where the remainder term is due to Assumption \ref{ass:unif}. The claim then follows from standard arguments. \hfill$\square$\\

\singlespacing 
\addcontentsline{toc}{section}{References}
\bibliography{bib}

\end{document}